\documentclass[fleqn]{2017SCGE}
\usepackage{mwe,tikz,ulem}
\usepackage[percent]{overpic}
\usepackage{graphicx}
\usepackage{xspace}
\usepackage[numbers]{natbib}
\usepackage[utf8x]{inputenc}

\usepackage[gen]{eurosym}
\usepackage{microtype}
\usepackage{color}
\usepackage{txfonts}
\usepackage{makeidx}
\usepackage[columns=1]{idxlayout}

\newcommand{\extp}{\textsl{eXTP}\xspace}

\newcommand{\lad}{LAD\xspace}
\newcommand{\wfm}{WFM\xspace}
\newcommand{\xmm}{\textsl{XMM-Newton}\xspace}

\newcommand{\apj}{ApJ\xspace}
\newcommand{\apjl}{ApJ\xspace}

\begin{document}
\ensubject{subject}

%%%%%%%%%%%%%%%%%%%%%%%%%%%%%%%%%%%%%%%%%%%%%%%%%%%%%%%
%%% Authors do not modify the information below
%%% ????????????????
%%% ??????????, ????????????{}, ???????????????????
%Letter to the Editor??Article%??????
\ArticleType{Article}%??Article
\SpecialTopic{SPECIAL TOPIC: }%???????
\Year{2017}
\Month{January}
\Vol{60}
\No{1}
\DOI{10.1007/s11432-016-0037-0}
\ArtNo{000000}
% \ReceiveDate{January 11, 2016}
% \AcceptDate{April 6, 2016}
%\OnlineDate{January 1, 2016}
%%%%%%%%%%%%%%%%%%%%%%%%%%%%%%%%%%%%%%%%%%%%%%%%%%%%%%%

%%%   \title{title}{title for citation}
\title{Physics and Astrophysics of Strong Magnetic Field systems with eXTP}{SM with eXTP}

\author[1,2]{Andrea Santangelo}{{andrea.santangelo@uni-tuebingen.de}}
\author[3]{Silvia Zane}{{s.zane@ucl.ac.uk}}
\author[4]{Hua Feng}{{hfeng@tsinghua.edu.cn}}
\author[5]{Renxin Xu}{{r.x.xu@pku.edu.cn}}
\author[1]{Victor Doroshenko}{doroshv@astro.uni-tuebingen.de}
\author[6]{\\Enrico Bozzo}{}
\author[9]{Ilaria Caiazzo}{}
\author[7,17,20]{Francesco Coti Zelati}{}
\author[17]{Paolo Esposito}{} 
\author[3]{Denis González-Caniulef}{}
\author[9]{\\Jeremy Heyl}{}
\author[10]{Daniela Huppenkothen}{}
\author[11]{Gianluca Israel}{}
\author[12]{Zhaosheng Li}{}
\author[13]{Lin Lin}{}
\author[8,15]{\\Roberto Mignani}{}
\author[16,17]{Nanda Rea}{}
\author[14]{Mauro Orlandini}{}
\author[18]{Roberto Taverna}{}
\author[19]{\\Hao Tong}{}
\author[3,18]{Roberto Turolla}{}
\author[21]{Cristina Baglio}{}
\author[26]{Federico Bernardini}{}
\author[28]{\\Niccolo' Bucciantini}{}
\author[30,31]{Marco Feroci}{}
\author[32]{Felix F\"urst}{}
\author[33]{Ersin G\"o\u{g}\"u\c{s}}{}
\author[2]{Can G\"{u}ng\"{o}r}{}
\author[1]{Long Ji}{}
\author[2]{\\Fangjun Lu}{}
\author[23,24]{Antonios Manousakis}{}
\author[8]{Sandro Mereghetti}{}
\author[22]{Romana Mikusincova}{}
\author[25]{Biswajit Paul}{}
\author[34]{\\Chanda Prescod-Weinstein}
\author[27]{George Younes}{}
\author[29]{Andrea Tiengo}{}
\author[2]{Yupeng Xu}{}
\author[17]{Anna Watts}{}
\author[2]{Shu Zhang}{}
\author[2]{\\Shuang-Nan Zhang}{}

% new authors after submission

%%% Author information for page head.

\AuthorMark{Santangelo A., Zane S., Feng. H., Xu R., et al.}%\authorcr????????

%%% Authors for citation.

\AuthorCitation{Santangelo A., Zane S., Feng. H., Xu R., et al.}

%%% Address.
%%%   \address[number]{Address, City {\rm Postcode}, Country}
%%\address[1]{Department, University, City, Postal code, Country;}
%%\address[2]{For example: Institute of Mechanics, Chinese Academy of Sciences, Beijing 100190, China}

\address[1]{Institut f\"{u}r Astronomie und Astrophysik, Eberhard Karls Universit\"{a}t, 72076 T\"{u}bingen, Germany}
\address[2]{Key Laboratory for Particle Astrophysics, Institute of High Energy Physics, Beijing 100049, China}
\address[3]{Mullard Space Science Laboratory, University College London, Holmbury St Mary, Dorking, Surrey, RH56NT, UK}
\address[4]{Department of Engineering Physics and Center for Astrophysics, Tsinghua University, Beijing 100084, China}
\address[5]{School of Physics, Peking University, Beijing 100871, China}
\address[6]{Department of Astronomy, University of Geneva, Chemin d'Ecogia 16, 1290 Versoix, Switzerland}
\address[7]{Universit\`a dell'Insubria, Via Valleggio 11, I-22100 Como, Italy}
\address[8]{INAF -- Istituto di Astrofisica Spaziale e Fisica Cosmica Milano, via E. Bassini 15, 20133, Milano, Italy}
\address[9]{Department of Physics and Astronomy, University of British
 Columbia, 6224 Agricultural Road, Vancouver, BC V6T 1Z1, Canada}
\address[10]{Center for Data Science, New York University, 726 Broadway, 7th Floor, New York, NY 10003, USA}
\address[11]{INAF -- Osservatorio Astronomico di Roma, Via Frascati 33, I-00040 Monteporzio Catone, Italy} 
\address[12]{Department of Physics, Xiangtan University, Xiangtan 411105, China}
\address[13]{Department of Astronomy, Beijing Normal University, Beijing 100875, China}
\address[14]{INAF -- Istituto di Astrofisica Spaziale e Fisica Cosmica Bologna, Via Gobetti 101, 40129 Bologna, Italy}
\address[15]{Janusz Gil Institute of Astronomy, University of Zielona G\'ora, Lubuska 2, 65-265, Zielona G\'ora, Poland}
\address[16]{Instituto de Ciencias del Espacio (ICE), CSIC-IEEC, 08193, Barcelona, Spain}
\address[17]{Anton Pannekoek Institute for Astronomy, University of Amsterdam, Postbus 94249,  NL-1090-GE Amsterdam, The Netherlands}
\address[18]{Department of Physics and Astronomy, University of Padova, via
Marzolo 8, 35131 Padova, Italy}
\address[19]{School of Physics and Electronic Engineering, Guangzhou University, Guangzhou 510006, China}
\address[20]{INAF -- Osservatorio Astronomico di Brera, Via Bianchi 46, I-23807 Merate (LC), Italy}
\address[21]{Physics Department, New York University Abu Dhabi, PO Box 129188, Abu Dhabi, UAE}
\address[22]{Institute of Theoretical Physics, Faculty of Mathematics and Physics, Charles University in Prague}
\address[23]{Centrum Astronomiczne im. M. Kopernika, Bartycka 18, 00-716, Warszawa, Poland}
\address[24]{Department of Physics, Sultan Qaboos University, 123 Muscat, Oman}
\address[25]{Raman Research Institute Sadashivanagar, C. V. Raman Avenue, Bangalore 560080, India}
\address[26]{NYU Abu Dhabi, Saadiyat Campus P.O. Box 129188 Abu Dhabi, UAE}
\address[27]{The George Washington University, Physics department, 725 21st St NW, Washington, DC 20052}
\address[28]{NAF Osservatorio di Arcetri, Largo Enrico Fermi 5, 50125  Firenze Italy}
\address[29]{Scuola Universitaria Superiore IUSS Pavia, Palazzo del Broletto, Piazza della Vittoria n.15 - 27100 Pavia, Italy}
\address[30]{INAF -- Istituto di Astrofisica e Planetologia Spaziali, Via Fosso del Cavaliere 100, I-00133 Roma, Italy} 
\address[31]{INFN -- Roma Tor Vergata, Via della Ricerca Scientifica 1, I-00133 Roma, Italy} 
\address[32]{European Space Astronomy Centre (ESA/ESAC), Science Operations Department, Villanueva de la Ca\~{n}ada, E-28692 Madrid, Spain} 
\address[33]{Sabanc\i~University, Faculty of Engineering and Natural Sciences, Orhanl\i -Tuzla, \.Istanbul 34956 Turkey}
\address[34]{University of Washington, Department of Astronomy, Box 351580, Seattle, USA}
%\contributions{}

%%% Abstract. 
\abstract{In this paper we present the science potential of the \textit{enhanced X-ray Timing and Polarimetry} (eXTP) mission for studies of strongly magnetized objects. We will focus on the physics and astrophysics of strongly magnetized objects, namely magnetars, accreting X-ray pulsars, and rotation powered pulsars. We also discuss the science potential of eXTP for QED studies. Developed by an international Consortium led by the Institute of High Energy Physics of the
Chinese Academy of Sciences, the eXTP mission is expected to be launched in the mid 2020s.}

%%% Keywords.


\maketitle

%\tableofcontents%?????

%%%%%%%%%%%%%%%%%%%%%%%%%%%%%%%%%%%%%%%%%%%%%%%%%%%%%%%
%%% The main text. ???????
%???????????????????\cref{fig1}
%\twocolumn\onecolumn
%%%%%%%%%%%%%%%%%%%%%%%%%%%%%%%%%%%%%%%%%%%%%%%%%%%%%%%
\begin{multicols}{2}
\section{The eXTP mission}\label{section1}
 
In this white paper we present the science potential of the \textit{enhanced
X-ray Timing and Polarimetry} (\extp) mission for studies of \textit{strongly
magnetized objects}. The scientific payload of \extp consists of four main
instruments: The Spectroscopic Focusing Array (SFA), the Polarimetry Focusing
array (PFA), the Large Area Detector (\lad) and the Wide Field Monitor (WFM).
A detailed description of \extp’s instrumentation, which includes all relevant operational parameters, can be
found in \cite{Zhang2018}, but we summarize it briefly here. 

The SFA is an array of nine identical X-ray telescopes covering the energy
range 0.5-10~keV, featuring a total effective area of 0.4~m$^2$ at 6~keV,
and close to $\sim0.5$~m$^2$ at 1~keV. The SFA angular resolution is
$\sim1$~arcmin and its sensitivity reaches $10^{-14}$erg~s$^{-1}$~cm$^{-2}$ for
an exposure time of $10^{4}$~ks. In the current baseline, the SFA focal plane
detectors are silicon-drift detectors (SDDs), which combine CCD-like spectral
resolution with very short dead time and a high time resolution of 10$\mu$s; they are therefore well-suited for studies
of the brightest cosmic X-ray sources at the shortest time scales.

The PFA consists of four identical telescopes, with angular resolution better
than $\sim30$~arcsec and total effective area of $\sim700$~cm$^2$ at 2~keV
(including the detector efficiency), featuring Gas Pixel Detectors (GPDs) to
allow polarization measurements in the X-rays. The PFA is sensitive in the
2-8 keV energy range, and features a time resolution better than 500$\mu$s. The
instrument reaches a minimum detectable polarization of 5\% in 100~ks for a
Crab-like source of flux $10^{-11}$erg~s$^{-1}$~cm$^{-2}$.

The \lad has a very large effective area of $\sim 3.4$~m$^2$ (at 6~keV),
obtained with non-imaging SDDs, active between 2 and 30~keV and collimated to a
field of view of 1~degree across. It features absolute timing accuracy
of $\sim 1\mu$s, and an energy resolution better than 260\,eV (at 6\,keV). The \lad and the
SFA together reach an unprecedented total effective area of $\sim4$~m$^{2}$.

The science payload is completed by the WFM, consisting of 6 coded-mask cameras
covering more than 3.2~sr of the sky at a sensitivity of 4~mCrab for an exposure time of
1~d in the 2 to 50~keV energy range, and for a typical sensitivity of 0.2~mCrab
combining 1~yr of observations outside the Galactic plane. The instrument will
feature an angular resolution of a few arcmin and will be endowed with an
energy resolution of about 300~eV.

Developed by an international Consortium led by the Institute of High Energy
Physics (IHEP) of the Chinese Academy of Sciences, the mission is expected to be
launched around 2025. In section~\S\ref{sec:2} we focus on the \extp science
potential for magnetar candidates, namely the anomalous X-ray pulsars 
(AXPs) and soft gamma ray repeaters
(SGRs). In section~\S\ref{AccrP} we discuss the
\extp capability for accreting and rotation powered pulsars. Finally, in
section~\S\ref{qed}, we present the expected \extp impact on QED studies.
\noindent

\section{The physics and astrophysics of AXPs/SGRs as magnetar candidates}\label{sec:2}

\subsection{Introduction}
\label{Intro_2}
\begin{figure}[H]
    \centering
    \includegraphics[scale=0.6]{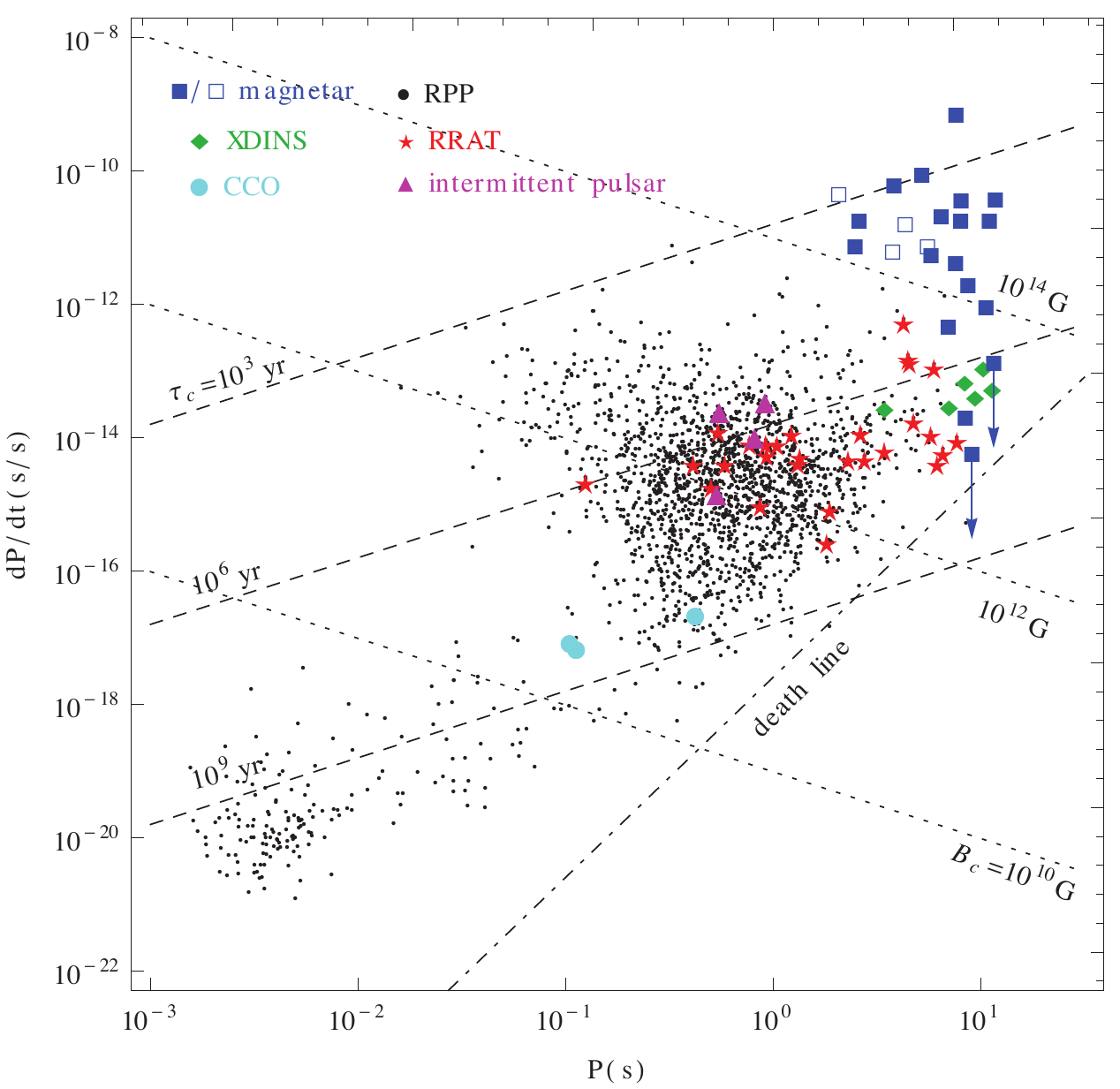}
    \caption{Period and period derivative diagram of neutron stars,
    including normal pulsars (black points), magnetars (blue squares, empty
    squares for radio loud magnetars), X-ray dim isolated neutron stars (green
    diamonds), central compact objects (light blue circles), rotating radio
    transients (red stars) and intermittent pulsars (magenta
    triangles)~\cite{ref5}.}
\label{fig:ppdot}
\end{figure}

Pulsars are rotating magnetized neutron stars (NSs), and are amongst the most
fascinating objects in astrophysics due to their potential as laboratory for physics under
extreme conditions. Their periodic modulation reflects their rotation period. 
They appear in many flavors and show
a rather different phenomenology depending on key parameters such as the spin
period and magnetic field intensity. The spin period and period derivative of
different types of pulsars are shown in Figure~\ref{fig:ppdot} together with
lines of constant B-field characteristic age $\tau$ (see e.g., \cite{ref1} for a review). Several classes
can clearly be identified, including the class of magnetars on the top right of
the figure.

Magnetars are NSs with extremely intense magnetic fields of the order of
B$\sim$10$^{14-15}$~G~\citep{ref2,ref3}, whose decay or recombination powers
their high energy emission in X-rays and even gamma rays.
Magnetars are thought to manifest as anomalous X-ray pulsars (AXPs)
and soft gamma repeaters (SGRs), with spin periods of $\sim (1-10)$~s and
period derivatives of $\sim(10^{-15}-10^{-10})~ {\rm s/s}$.

From X-ray observations, nearly thirty magnetar candidates have been identified
so far\footnote{http://www.physics.mcgill.ca/$\sim$pulsar/magnetar/main.html}.

\begin{figure*}[ht!]
    \centering
    \includegraphics[width=0.35\textwidth, angle =90]{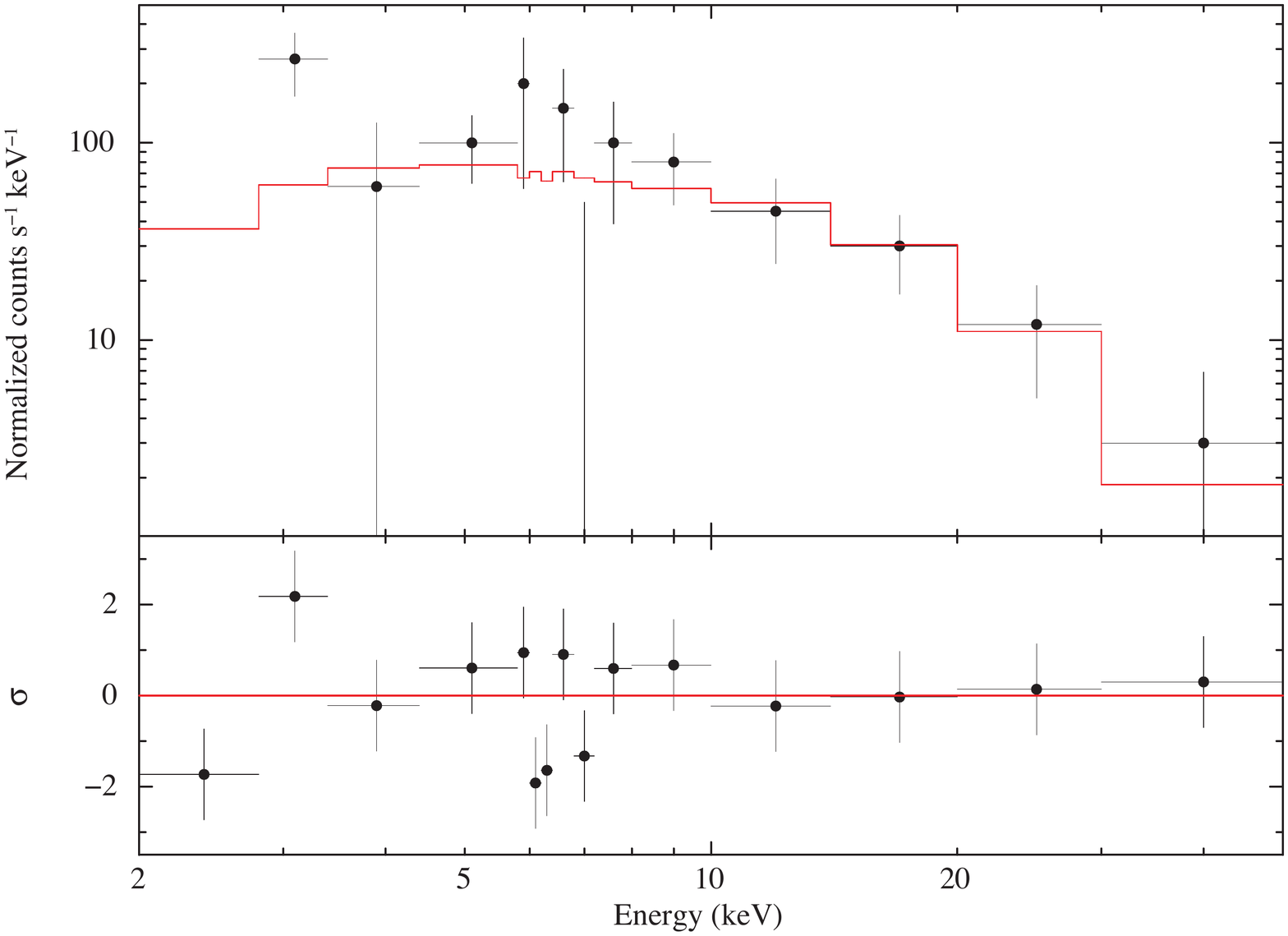}
    \includegraphics[width=0.5\textwidth]{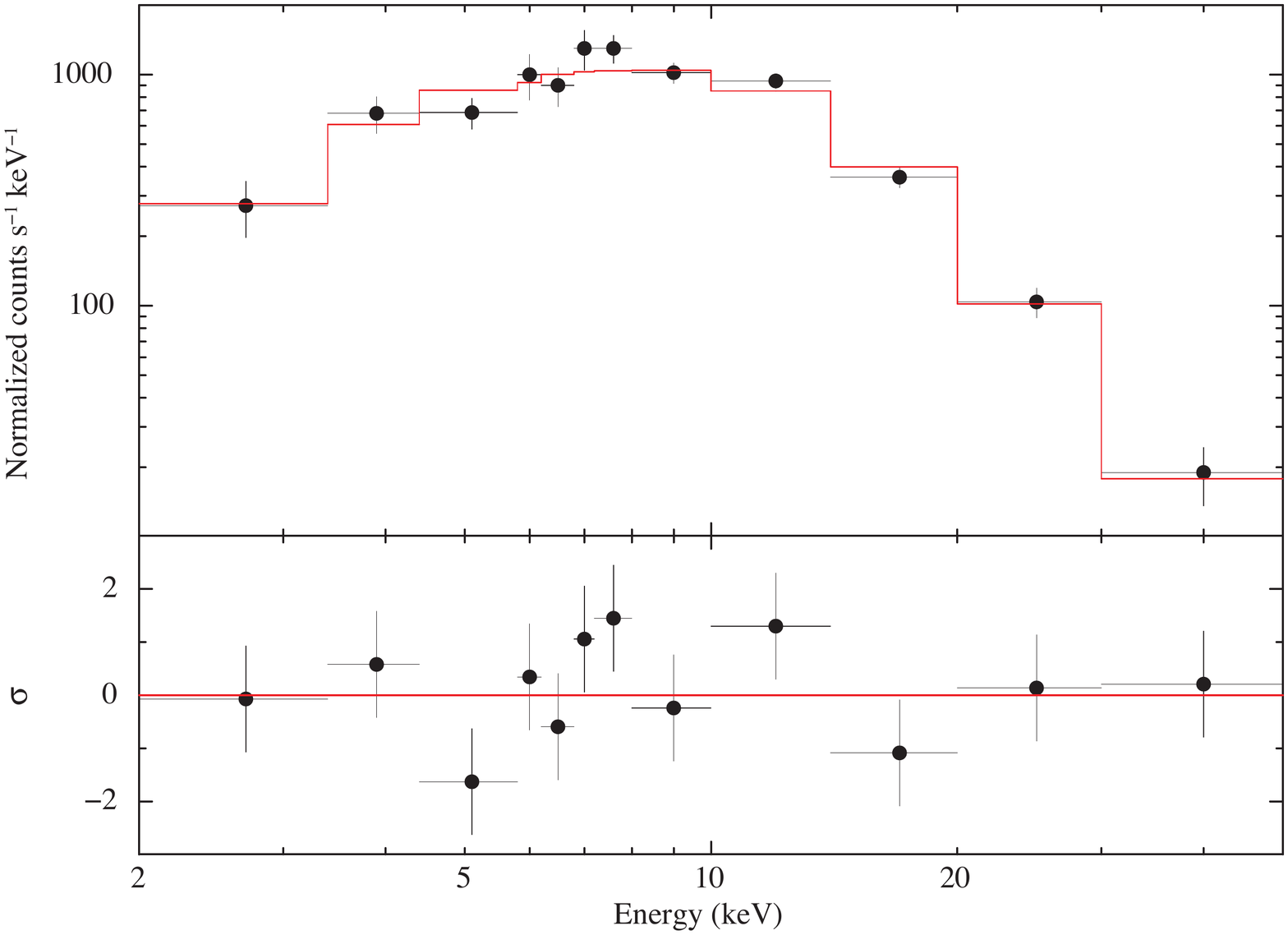}
    \caption{Left: Simulated WFM spectrum of a magnetar burst (black points)
    for a 0.05~s burst duration, assuming a flux of 10~Crab (left) and~150 Crab
    (right), and a double blackbody spectrum.} 
    \label{fig:burstline1}
\end{figure*}

AXPs/SGRs feature energetic events not seen in most pulsars. They emit
\textit{short bursts} of X-rays and soft $\gamma$-rays, which often reach
super-Eddington luminosities. More rarely, they also exhibit
\textit{intermediate} (IFs) and \textit{giant flares} (GFs), involving the
release of up to $10^{46}~ {\rm erg}$ in less than half a second. AXPs/SGRs are
also sources of persistent pulsed X-ray emission with typical luminosity of
$\sim10^{35}~{\rm erg/s}$.

They are ideal targets for \extp, since the broad band, high sensitivity,
polarimetric and monitoring capability of the mission will lead to the prompt
discovery of outbursts from known and new sources, thus enabling deep studies of
magnetar phenomenology on different time scales.

In fact, \extp will have an unprecedented capability to study their bursts,
outbursts (\S\ref{burst}), and their timing behaviour (\S\ref{timing}),
including glitches and precession. \extp observations will allow us to measure braking indices
in a more reliable way, and to perform asteroseismology.
Phase dependent absorption spectral features could also be 
observed systematically by \extp, enabling investigations of the intensity and topology of
the magnetic field configuration in the vicinity of the NS (\S\ref{lines}).
Furthermore, a combined study of spectra, timing and polarization may result in
key detections of magnetars in binary systems. All of these are essential studies
to unveil the nature of magnetars and to answer long-standing open questions in
magnetospheric and fundamental physics. (\S\ref{binary}).

\subsection{Bursts and Outbursts}
\label{burst}

Magnetars are highly variable X-ray sources, with the flux changing by as much
as a factor of thousand. Variability occurs on different times scales, e.g.
short bursts ($< 1$~s), intermediate flares ($1-40$~s), giant flares
($\sim$500~s), and outbursts that can last several weeks to years.

Indeed, most magnetar candidates were discovered thanks to their bursting
and transient X-ray emission by previous wide field monitors such as the
Fermi's GBM and the Burst Alert Monitor on Swift. Giant flares are very rare
events and have not been observed to repeat in a single source. Both bursts and
intermediate flares can recur on a time scale of a few hours.

\subsubsection{Burst Emission: spectral and polarization properties}

Magnetar bursts show a variety of spectral and timing behaviors \cite{Turolla2015}. 

The emission process is generally assumed to be initiated by rapid
rearrangements of the magnetic field, possibly involving either induced or
spontaneous magnetic reconnection. This accelerates charged particles with
ensuing gamma-ray emission, since the rapid acceleration of electrons in a
strongly curved field leads to a cascade of pair creation and
gamma-rays~\citep{ref6}. However, fully self-consistent models of the emission
resulting from the various proposed trigger mechanisms do not yet exist.
Considering their variability time scales and the WFM sky coverage, \extp
may discover a new candidate magnetar every year, triggering follow-up
observations with the \extp narrow field instruments and other facilities. These
follow-up observations of bursts and IFs are crucial to inferring the NS's interior
properties (see \S\ref{seismo}) and to search for proton cyclotron resonance
scattering features (CRSF) in the X-ray spectrum (e.g.,~\cite{ref7}, see
\S\ref{lines}).

\begin{figure}[H]
    \centering
    \includegraphics[width=0.45\textwidth]{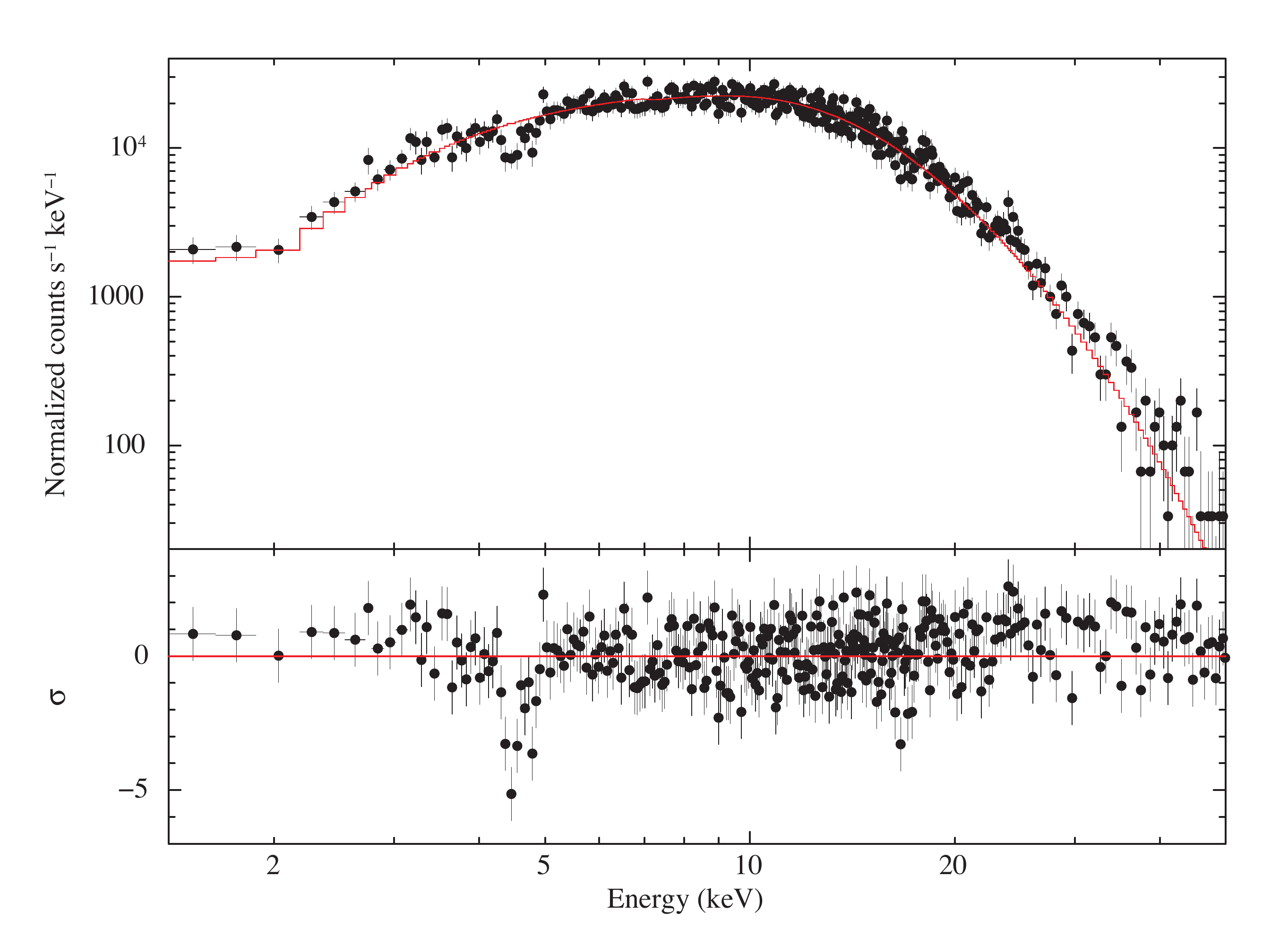}
    \caption{Simulated \lad spectrum of a magnetar burst (black points) for a
    0.05 s burst duration, assuming a flux of 10 Crab, a double blackbody
    spectrum, and an absorption feature at 4.5~keV (200~eV EW). The red line in
    the top panel shows the best-fit model obtained assuming a double 
blackbody spectrum. The lower panel shows the residuals with respect to the
    best-fit model, when the CRSF is not included.}
    \label{fig:burstline2}
\end{figure}

Assuming a 0.05~s burst duration, a peak flux of 150~Crab, and a spectrum
consisting of two absorbed blackbodies with $kT_1=5.4$~keV, $kT_2=10.2$~keV and $
N_H=2.1 \times10^{22}$~cm$^{-2}$ (as in the event observed from SGR 
1900+14,~\cite{ref8}),
we obtain a WFM count-rate of $(1.40\pm0.05)\times10^4$ cts~s$^{-1}$, showing
that such a burst will be detected with high significance (see
Fig~\ref{fig:burstline1}).

The fit of a simulated \lad spectrum of a much fainter burst of 10~Crab with a
proton cyclotron resonant scattering feature (CRSF) at 4.5~keV, and equivalent
width of 200~eV (shown in Fig~\ref{fig:burstline2}), shows that these
absorption features can be detected with a high significance (at more than
8$\sigma$) within a single burst. These features will be discussed in 
detail in \S\ref{lines}.

\begin{figure*}[ht!]
    \centering
    \includegraphics[width=0.9\textwidth]{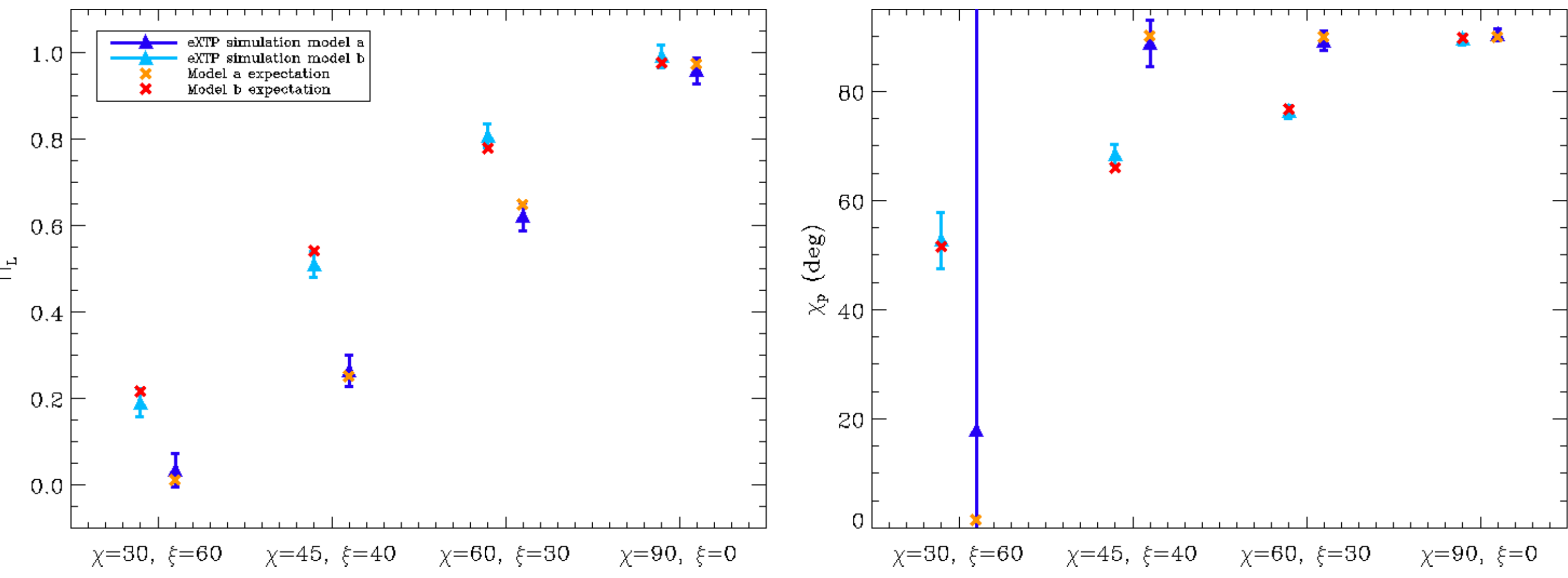}
    \caption{
Simulated polarization fraction (left panel) and angle (right panel) as 
measured by the \extp polarimeter for an intermediate burst computed 
according to model a (orange crosses) and model b (red crosses) 
of~\cite{{Taverna2017}}; 
the simulated \extp response is shown by the light-blue and dark-blue 
triangles, respectively, with 1$\sigma$ error bars. An exposure time of 
1.737 s (i.e. the total duration of the burst) is assumed, together with a 
$1$--$10$ keV flux of $4.68\times10^{-7}$ erg/cm$^{2}$/s. The simulations 
are performed for different values of the geometrical angles $\chi$ and 
$\xi$, which represent the inclination of the line-of-sight and of the 
magnetic axis with respect to the rotation axis, respectively.}
\label{fig:burstpolarization}
\end{figure*}

In order to test \extp's ability to measure polarization 
properties of magnetar bursts, we have produced several simulated data sets, using for 
the burst the model recently developed by~\cite{Turolla2015}. The burst spectral and
polarization properties were computed within the fireball scenario
developed by~\cite{Taverna2017}, both in the case in which the pair plasma fills the entire
toroidal region limited by a set of closed field lines (model a), and in the case where the
emitting region is confined to a portion of the torus-like volume (model b).
Simulations were performed for a typical intermediate flare, like those emitted
by SGR 1900+14 during the ``burst forest'' of 2006~\cite{ref8}, with duration
$\sim 1.7$ s and $1$--$10$ keV flux $\sim 5\times 10^{-7}\ {\rm erg/cm^2/s}$.
Results are shown in Fig~\ref{fig:burstpolarization}.

Both polarization fraction and angle measurements recover, within 
$1\sigma$ uncertainties, values used in the simulations. It 
should be noted that although the errors on the polarization fraction are 
more or less the same for all the geometrical configurations considered, 
those on the polarization angle increase by decreasing the corresponding 
degree of polarization.  Instrumental effects only dominate the polarization angle measurements for $P \lesssim 20\%$.

\subsubsection{Outbursts' Decay}

The largest amount of physical information on magnetars is gathered during
outbursts, when the emission is enhanced at all wavelengths and the variability
is rapid, challenging our present knowledge of heating mechanisms and NS
thermal inertia. During outbursts, the X-ray flux of the source suddenly
increases by a factor $\sim (10-1000)$ over the persistent level in a few
hours. During the active phase, which lasts $\sim1$~year, the flux declines, the
spectrum softens and the pulse profile simplifies. Since such outbursts are
likely produced by deformations of the neutron star crust, which, in turn,
produce a rearrangement of the external magnetic field over a large
scale~\cite{ref9}, dedicated observational campaigns of transient sources offer
the best opportunity to monitor the evolution of the magnetosphere at different
luminosities and at high spectral resolution, over several months/years.

The \lad and SFA can monitor the light curve, pulse profiles and spectra over a
broad energy range, from a few hours after the onset of the outburst to its
decay to the quiescent state, at a level of precision unattainable by
NuSTAR~\citep{ref10}.

The combination of \extp instruments will enable us to study the correlation
between variabilities in at least two bands, and will reveal the slope
turn-over in the soft-to-hard emission, generating a deep
understanding of the magnetic field and current distributions. \extp
phase-resolved spectroscopy will allow us to track the variability of 
the magnetospheric configuration during the outburst decay, following the variations
in the power-law spectral component, and to search for CRSF variability. 
For objects that reach surface temperatures as high as $\sim1$~keV, phase-resolved
spectroscopy will also allow us to map the region of the NS surface heated 
during the early phases of the outburst. 

A unique and unprecedented characteristic of \extp is that it will perform 
timing and polarization measurements simultaneously. The change of the 
angle and degree of polarization during the outburst, combined with measurements 
of flux and spectral evolution, will yield a unique and novel data set to
constrain the magnetospheric topology and remove existing degeneracies in
proposed models~\cite{ref11}.

Data will also allow us to compare detailed predictions from the twisted
magnetosphere and thermal relaxation models: different polarization 
signatures are expected, depending on which effect dominates, and quantitative theoretical
studies are presently in progress.  Furthermore, polarimetric measurements of
the thermal X-ray component of transient magnetars during an outburst decay will
provide a crucial tool to prove magnetospheric untwisting and in turn to
test the magnetar scenario for the post-outburst relaxation epoque, as shown later in
sec. 4.2, and in \cite{refG17,refG17b}.

\subsection{Timing Properties}
\label{timing}

Thanks to the large field of view of the WFM, a large fraction of
the Galactic Plane (where most magnetar candidates have been discovered) 
will be covered
during most \extp pointings. This gives us the possibility of monitoring the spin
period of several magnetars simultaneously, to obtain phase-connected timing
solutions. 
For X-ray fluxes of a few mCrab, the WFM will accurately measure magnetar
pulsations in a few tens of ks.
We note that all of the persistently bright magnetars are radio quiet and 
hence X-ray timing is the only way of measuring their spin evolution.
Via the monitoring of the spins, we will be able to detect glitches, 
possibly reveal precession, and accurately measure braking indices, as
well as other timing irregularities. We can investigate whether these events
anticipate bursts or giant flares. The dramatic increase in the effective area
provided by \lad/SFA in comparison to previously flown missions is also likely to reveal new timing features. The observed timing
behaviors could provide valuable information about the physical mechanism of
energy release and the radiative process.

\subsubsection{Glitches}

\begin{figure}[H]
    \centering
    \includegraphics[width=0.45\textwidth]{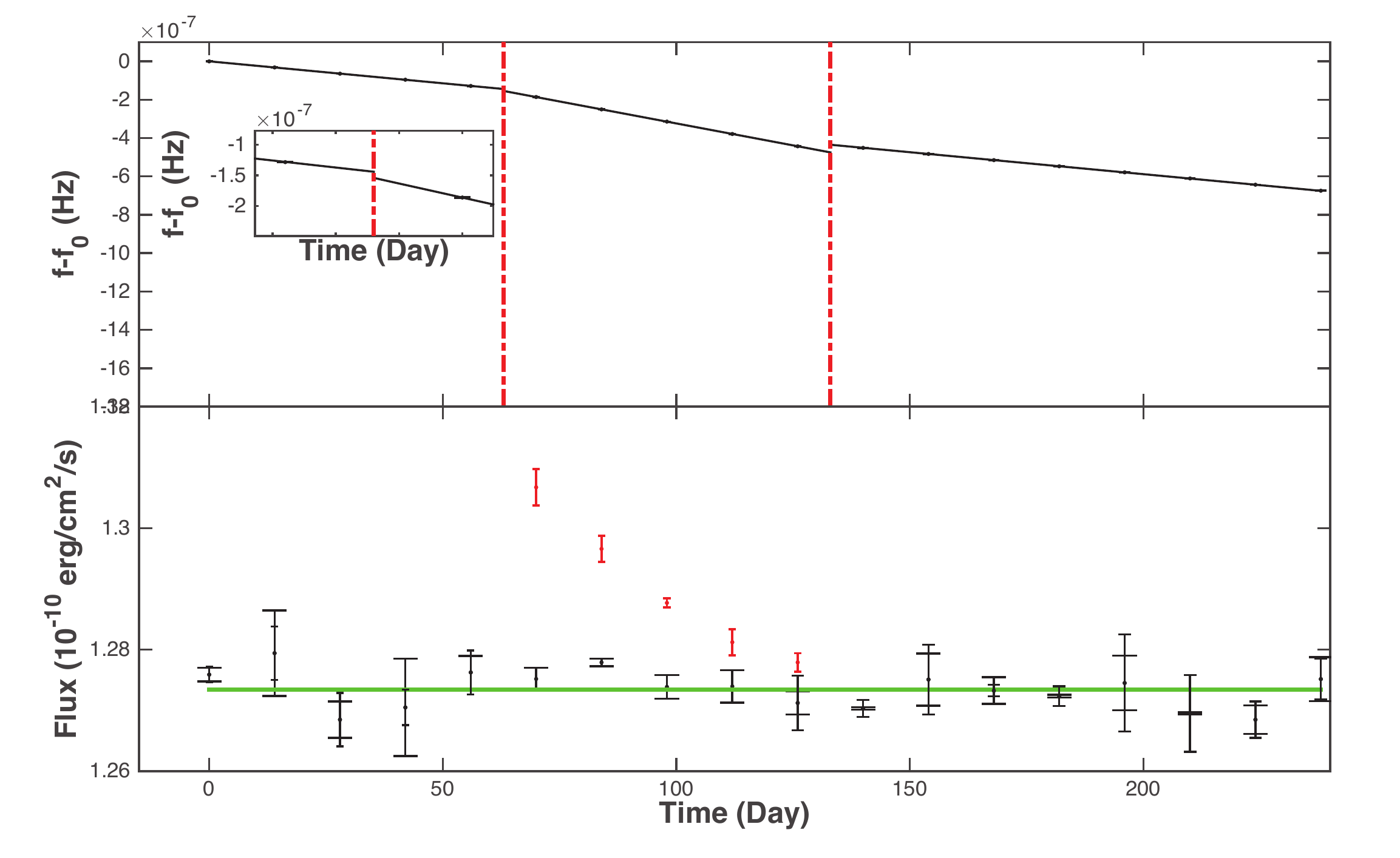} 
    \caption{\extp simulation of a small glitch with/without X-ray outburst.
    The input spectrum is a combination of a blackbody and a power law,
    absorbed at lower energies. The hydrogen column density, the blackbody temperature and its normalization are
    $0.95\times10^{22}~{\rm cm^{-2}}$, $0.395~{\rm keV}$ and $0.1558$,
    respectively. Points indicate observations of 2 ks spaced by intervals of 2 weeks, hence the total exposure time is
    $3.34\times10^{4}~\rm{s}$. In the bottom panel, the red data show the X-ray
    outburst during a small glitch, while the black data represent the
    persistent emission.}
    \label{fig:glitch}
\end{figure}

Magnetars (AXPs and SGRs) show peculiar timing behaviors, with frequent  
glitches and occasional anti-glitches~\citep{ref12}. The fractional change 
in the spin frequency is
typically $\Delta{\nu}/{\nu} \sim (10^{-7}-10^{-4})$ for AXPs/SGRs, similar to
that observed during large glitches in radio pulsars. The lack of smaller
glitches reflects the limited observational capability of X-ray
facilities, due to the combined effect of the larger timing noise of 
magnetars compared to radio pulsars and the lower cadence of timing 
observations \citep{ref12a, ref12b}. 
The glitches are usually (but not always) accompanied by X-ray flux
enhancement. Notably, SGR 1900+14 and 1E 2259+586 experienced events in which
the spin frequency jumps to a value smaller than that predicted by the observed
spin-down rate (with negative $\Delta \nu$, as for anti-glitches). Within 
the
conventional magnetar model, the problem of fully understanding glitches 
(with or
without X-ray enhancement) is still challenging. Several explanations have been
proposed, including collision with small solid bodies~\citep{ref13}, emission
of enhanced particle winds~\citep{ref14}, and starquakes in a solid strangeon
star~\cite{ref15,ref16}. In order to test these models, it is necessary to
detect glitches at the extremes of their amplitude, i.e. either very large 
or very small. As shown in Fig.~\ref{fig:glitch}, simulations indicate that in a
source as 4U 0142+61, \extp could detect a glitch/antiglitch with amplitude
$\Delta \nu/\nu\sim10^{-8}~\rm {s/s}$ in about 250\,d assuming 2\,ks 
monitoring observations with a two week
cadence. Furthermore, \extp observations will allow us to detect or rule 
out X-ray
enhancement during a small glitch (see the bottom panel of
Fig.~\ref{fig:glitch}). This is extremely important since the smaller
glitches/antiglitches could put tighter constraints on the various timing
anomaly mechanisms if no faint X-ray outbursts occur.

\subsubsection{Precession}

\begin{figure}[H]
    \centering
    \includegraphics[width=0.45\textwidth]{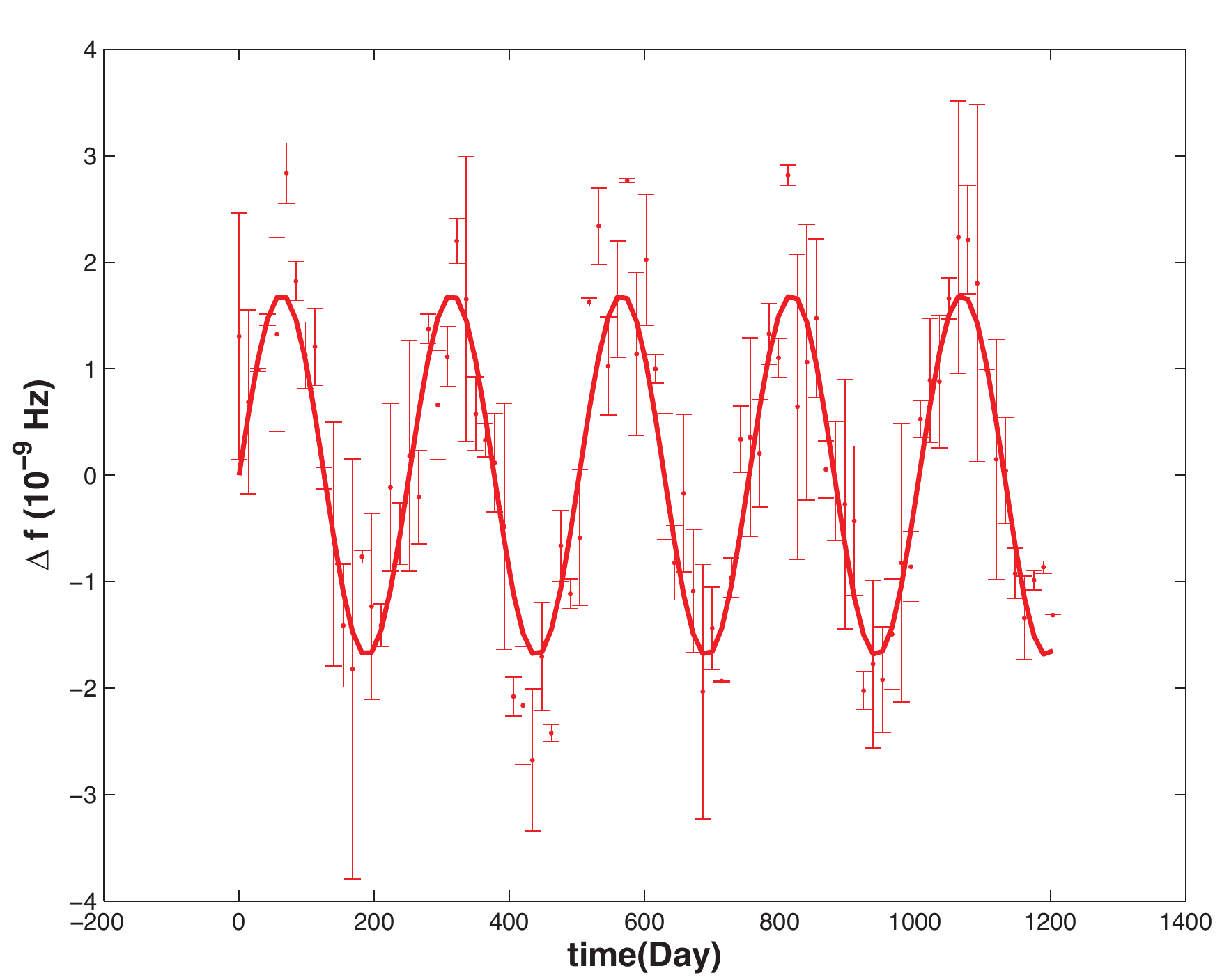}
    \caption{\extp simulation of free precession in SGR 1900+14. The input
    spectrum is a blackbody plus a power law, with low energy absorption. The
    hydrogen column density, the blackbody temperature and its normalization
    for SGR 1900+14 are $1.6\times10^{22}~{\rm cm^{-2}}$, 0.5 and 7.1\,keV,
    respectively. The sinusoidal modulation of the free precession is assumed.
   Points indicate observations of 2 ks spaced by intervals of 2 weeks, hence the total exposure time is
    $1.15\times10^{5}~\rm{s}$. }
    \label{fig:free}
\end{figure}

In the magnetar model it is expected that strong dipolar and multipolar
magnetic fields could induce large stresses on the star, in turn inducing
significant distortions of the star's shape. As a consequence, free 
precession
of the NS may manifest as a modulation of the spin frequency~\cite{ref17}. Such a
modulation has not yet been firmly observed. Nevertheless, possible evidence
for magnetar precession has been suggested from the analysis of X-ray phase
modulations~\cite{ref18}. The time scale of free precession is inversely
proportional to the square of the internal magnetic field $B_{\rm in}$,
\begin{equation} \tau_{\rm pr}={2\pi\over \epsilon \Omega}\simeq
2\times 10^{-2}\left ({B_{\rm in}\over 100B_{\rm QED}} 
\right)^{-2}\left ({P\over 6{\rm
s}} \right )~{\rm yr},
\end{equation}
where $B_{\rm QED}=4.414\times 10^{13}$ G is the quantum critical field. 
A firm detection of precession would have several implications 
for superconductivity in magnetar interiors~\cite{ref18a, ref18b}. On 
the other hand, if some of the alternative models that suggest AXPs and SGRs host normal magnetic
fields with $B_{\rm in}\sim 10^{12}$ G, are correct, precession might not
necessarily occur, either because the time-scale would be too long or because the
deviation from spherical symmetry in the compact star (with mass $\sim
M_\odot$) would be negligible, e.g.,~\cite{ref19}. \extp will search for
precession in AXPs/SGRs, ultimately testing the magnetar model and
constraining the star's parameters. As shown in Figure~\ref{fig:free}, 
systematic temporal monitoring with \extp will allow us to detect the amplitude of
free precession down to a level of $\Delta \nu\sim10^{-9}~ {\rm Hz}$, in a
magnetar like SGR 1900+14. 

\subsubsection{$\ddot{\nu}$ and braking index of magnetars}
\begin{figure}[H]
    \centering
    \includegraphics[width=0.5\textwidth]{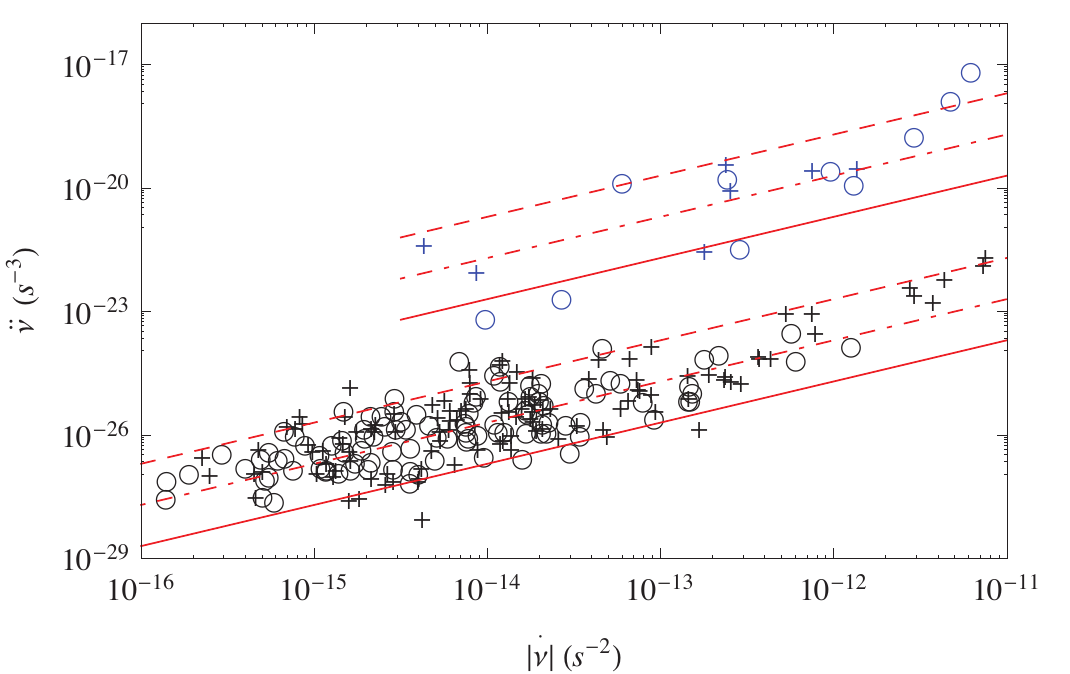}
    \caption{$\ddot{\nu}$ versus $\dot{\nu}$ for pulsars and magnetars. Black
    `+' stands for pulsars with postive $\ddot{\nu}$ and black `$\circ$' for
    negative $\ddot{\nu}$. Blue `+' and `$\circ$' are for magnetars. The lines
    are model calculations with different fluctuation amplitude. Updated from
    Figure 9 in~\cite{ref22}.}
    \label{fignu2dotmodel}
\end{figure}

The braking index $n$ of pulsars is defined as
\begin{equation}\label{braking_index}
n = \frac{\nu \ddot{\nu}}{\dot{\nu}^2},
\end{equation}
where $\dot{\nu}$, and $\ddot{\nu}$ are the pulse frequency
derivative, and frequency second derivative, respectively. The braking index
reflects the physics of the pulsar spin-down torque, through its effect on the
angular velocity. 
To date, 8 pulsars have a measured braking
index~\citep{ref20}, and more than 300 pulsars have a second derivative of the
frequency reported~\citep{ref21}. However, the frequency second derivatives of
these pulsars may be dominated by timing noise. For 17 magnetars the second
derivative of the frequency has been reported. Among them, 7 show a positive
$\ddot{\nu}$, while the other 10 systems have a negative $\ddot{\nu}$. Furthermore,
the frequency second derivatives of pulsars and magnetars are correlated with
their spin-down rate, as shown in Fig.~\ref{fignu2dotmodel}. There are at least
three causes that may contribute to the $\ddot{\nu}$ observed in 
magnetars. They are, in 
order of decreasing importance:
\begin{enumerate}
    \item{}Fluctuations in the magnetosphere. This process may in fact dominate
    the currently observed $\ddot{\nu}$ in magnetars and pulsars and also 
    explain the observed correlation (see again Fig.~\ref{fignu2dotmodel}). The
    corresponding second derivative is given by ~\citep{ref22} as

\begin{equation}
    \label{nu2dot_periodic_case}
    |\ddot{\nu}| = 2\pi \delta \frac{|\dot{\nu}|}{T},
    \end{equation} 
where $\delta$ and $T$ are the typical amplitude and timescale of the
magnetospheric fluctuations.

\item{} A decreasing spin-down rate, due to a decreasing particle wind in the
magnetosphere~\citep{ref23,ref24}. This process naturally explains a
corresponding positive $\ddot{\nu}$. However, the predicted magnitude is about
an order of magnitude smaller than the actual observed values.

    \item{} A braking index about one. 
Again, the corresponding values of
    $\ddot{\nu}$ (from equation~\ref{braking_index}) are several orders of
    magnitude smaller than the current observed values. 
\end{enumerate}

Since the quiescent states of transient magnetars are less noisy, more accurate
timing during the quiescent state may reveal the $\ddot{\nu}$ corresponding to
a decreasing spin-down rate/wind or, at least for some sources, may allow us to
test the determination of the braking index. \extp's \lad has larger
detection area, and thus is more effective in timing of magnetars in the X-ray
band. This will permit us to conduct regular monitoring of spin frequency changes
for a sample of sources using comparatively short exposures, and thus constrain the
braking index in more sources. 

\subsubsection{Asteroseismology}
\label{seismo}
In the era of CoRoT and Kepler, seismology has become firmly established as a
precision technique for the study of stellar interiors. The detection of
seismic vibrations in NSs was one of RXTE's most exciting discoveries, as 
it allows a unique, direct view of the densest bulk matter in the Universe.
Vibrations, detectable as quasi-periodic oscillations (QPOs) in the hard 
X-ray emission, were found in the tails
of giant flares from two magnetars~\citep{ref25,ref25a, ref25b,ref26}. Other 
candidate QPOs have 
since been
discovered during storms of short, low fluence bursts from several
magnetars~\citep{ref27,ref28,ref29}

Seismic vibrations from magnetars can in principle tightly constrain both the
interior magnetic field strength (which is hard to measure directly) and the
EOS~\citep{ref30,ref31}. They can also go beyond the EOS, constraining the
non-isotropic components of the stress tensor of supranuclear density material.
Seismic oscillation models now include the effects of the strong magnetic field
that couples crust to core in
magnetars~\citep{ref32,ref33,ref34,ref35,ref36,ref37}, superfluidity,
superconductivity and crust composition~\citep{ref38,ref39,ref40,ref41}. More
sophisticated models are under development, and much of the current effort is
focused on numerical implementation.

\begin{figure*}[ht!]
    \centering
    \includegraphics[width=1.\textwidth]{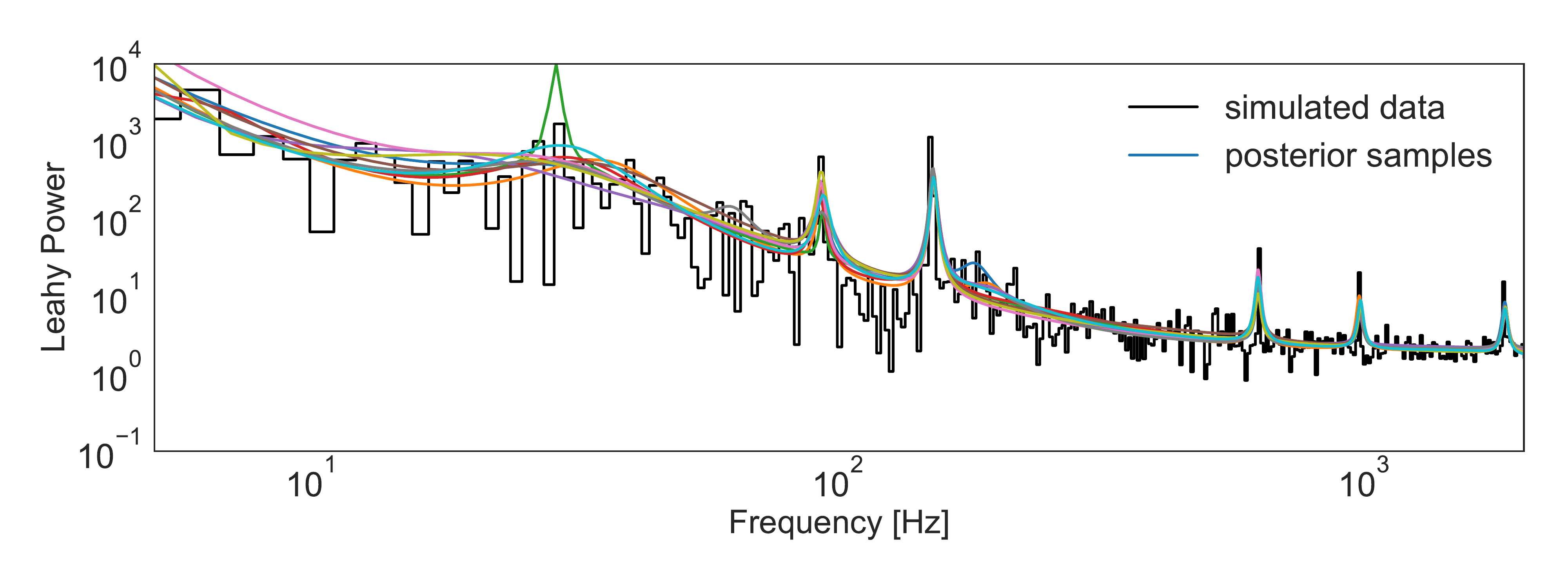}
    \caption{ Black: simulated \lad \extp data of the lightcurve of the IF
    observed from SGR~1900$+$14 in 2006, with quasi-periodic oscillations at 30
    Hz, 92 Hz, 150 Hz, 625 Hz, 976 Hz, and 1840 Hz, at a fractional rms
    amplitude of $50\%$ (with respect to that observed from the GF of
    SGR~1806$-$20). Overplotted in color are ten realizations from the
    posterior distribution of a mixture model of Lorentzian components, clearly
    identifying all but the lowest-frequency QPO present in the data.}
\label{fig:qpoIF}
\end{figure*}

With its large collecting area and excellent timing resolution, \extp would be
ideally placed to find QPOs in magnetar GFs, should one occur during the
mission lifetime. However, they are rare events, occurring only once every
$\sim$~10 years. What makes \extp unique is that it will be sensitive to QPOs of
less bright (but still $\sim 10^3$~Crab) IFs, which occur much more frequently. IFs have
similar peak fluxes as the tails of the GFs, but are too brief ($\sim 1$~s) to
permit detection of similar QPOs with current instrumentation.
Figure~\ref{fig:qpoIF} shows a simulation of the power-spectrum of the IF observed
from SGR~1900$+$14 in 2006, as seen by \extp. 
We artificially added to the
simulated lightcurve several QPOs at frequencies similar to those observed
in magnetar GFs (i.e. all those detected in the GF of SGR~1806$-$20 plus few
additional ones), using a fractional rms amplitude of 2, 1, and 0.5 in units of
the rms amplitude observed in the GF. Simulations indicate that the \lad onboard
\extp will be able to detect QPOs in IFs down to an amplitude of 0.5 of that
observed in the GFs, using only data from a single IF, without the need to
stack several events. Although the reported simulation is valid for events
detected on-axis, we expect that -- depending on the flux level of the IF -- it
will be possible to perform a detection at a similar level in bursts detected
off-axis.  Both IFs and GFs are in fact more likely to be observed off-axis due
to their unpredictability. Depending on the collimator properties, we
 expect a limited reduction in sensitivity, of the order of $\sim30$\%, for
off-axis detections (detailed simulations for off-axis events are underway and
will be reported in due course). In addition, \extp can make pointed
observations of bursting magnetars when they enter an active phase (when
bursting frequency increases) to detect many short, low fluence bursts.
Applying the techniques of \cite{ref27}, and stacking bursts together, it will
be possible to detect vibrations during magnetar burst storms.

\subsection{Spectral lines}
\label{lines}
Tiengo et al. (2013)~\cite{ref42} reported the discovery of a
phase-variable absorption feature in the X-ray spectrum of SGR 0418+5729. The
feature was best detected in a 67~ks \xmm observation performed on August 2009,
when the source flux was still particularly bright ($5\times 10^{12}$ 
erg~cm$^2$~s$^{-1}$ in
the 2-10~keV band). It has also been detected in RXTE and Swift data obtained
in the first two months after the outburst onset. The line energy lies in the
range 1-5 keV and changes sharply with the pulse phase, by a factor of $\sim
5$ in one-tenth of a cycle.

The most viable interpretation is in terms of a proton CRSF produced in a
baryon-rich magnetic structure (a small loop) localised close to the NS
surface~\citep{ref43}. The energy of proton CRSF scales linearly with the magnetic field as
$E_p \sim 0.44 B_{14}$~keV (assuming a standard NS EOS) and thus, for fields in
the magnetar range, they are expected to be observed below $\sim 6$~keV.

If the proton CRSF interpretation is correct, the presence of the line 
probes the complex topology of the magnetar magnetic field close to the 
surface and provides evidence of multipolar components of the star's 
magnetic field. This strengthens the idea that magnetar bursting behavior 
is dictated by high magnetic fields confined near the surface/inside the 
crust rather than by the large scale magnetic dipolar field. If protons, 
e.g. contained in a rising flux tube close to the surface, are responsible 
for the line, the local value of the magnetic field within the 
baryon-loaded structure is close to $10^{15}$~G.

Similar lines have been discovered in at least one further magnetar,
J1822.3-1606 \cite{ref44}, and at lower energies in some members of the 
XDINS class (a family often thought to be evolutionarily related to 
magnetars \cite{ref44a,ref44b}).

The \extp instruments will perform a systematic search and study of spectral
lines in magnetars (and of related families of NSs, as the XDINSs), ultimately
constraining the geometry of magnetars' surface magnetic field by performing
spin-phase resolved spectral analysis of proton cyclotron lines.

This study can be extended to magnetars in outburst, so as to follow 
magnetic twisting and relaxation during and after the outburst. About one new magnetar in outburst per year is expected to be detected during \extp's
life time, based on current statistics.

\begin{figure*}[ht!]
    \centering \includegraphics[width=0.4\textwidth]{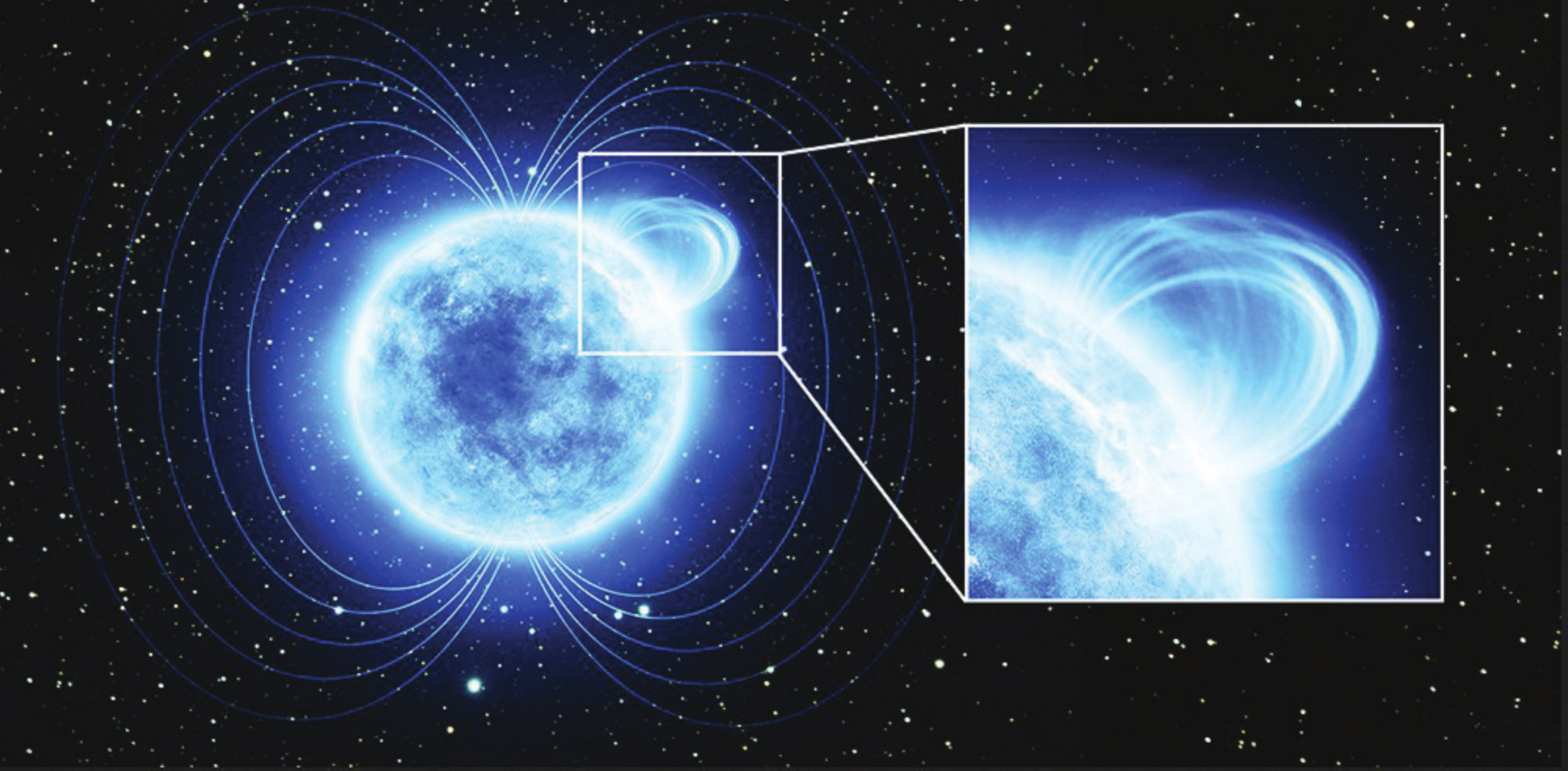}
    \includegraphics[width=0.4\textwidth]{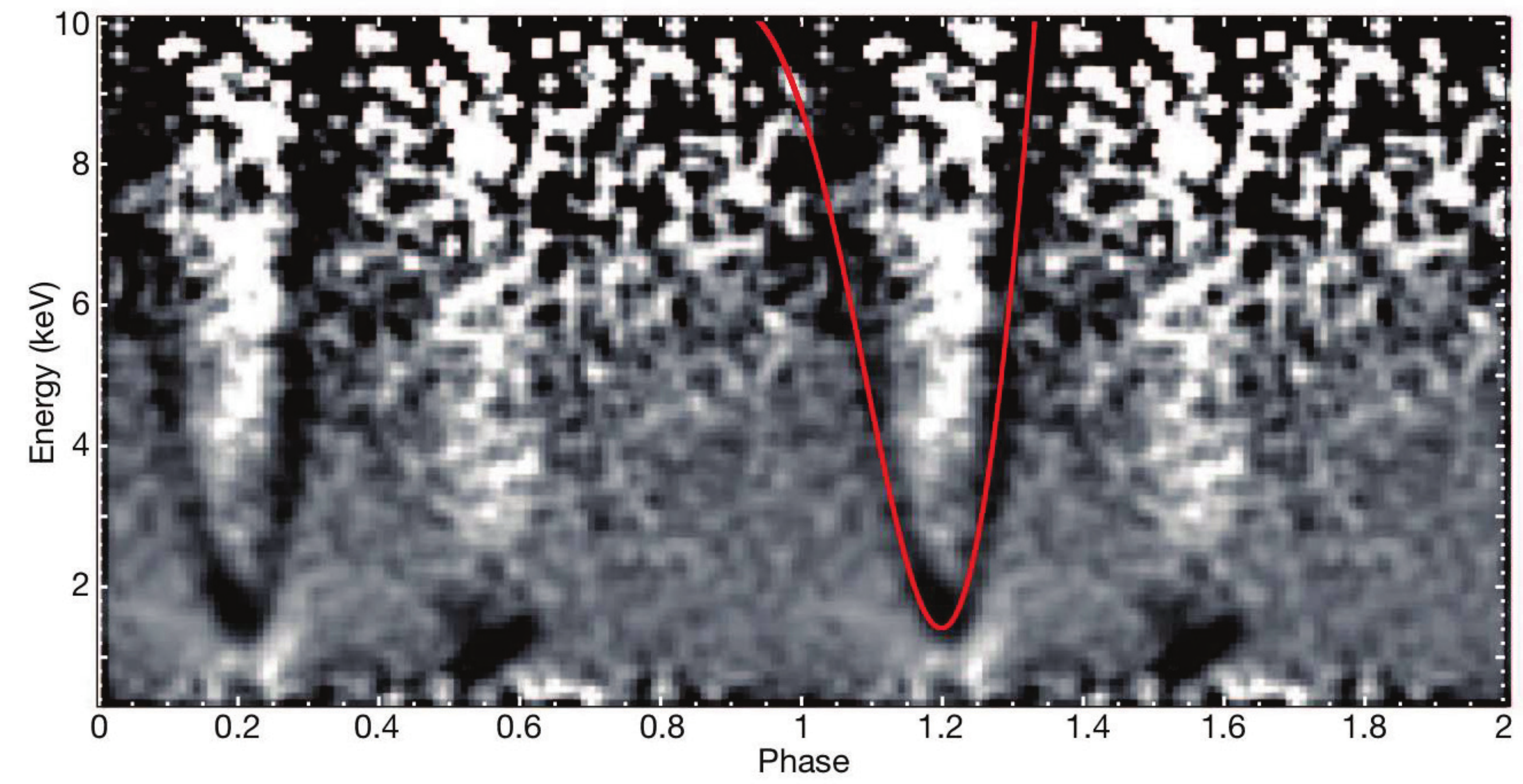}
    \includegraphics[width=0.4\textwidth]{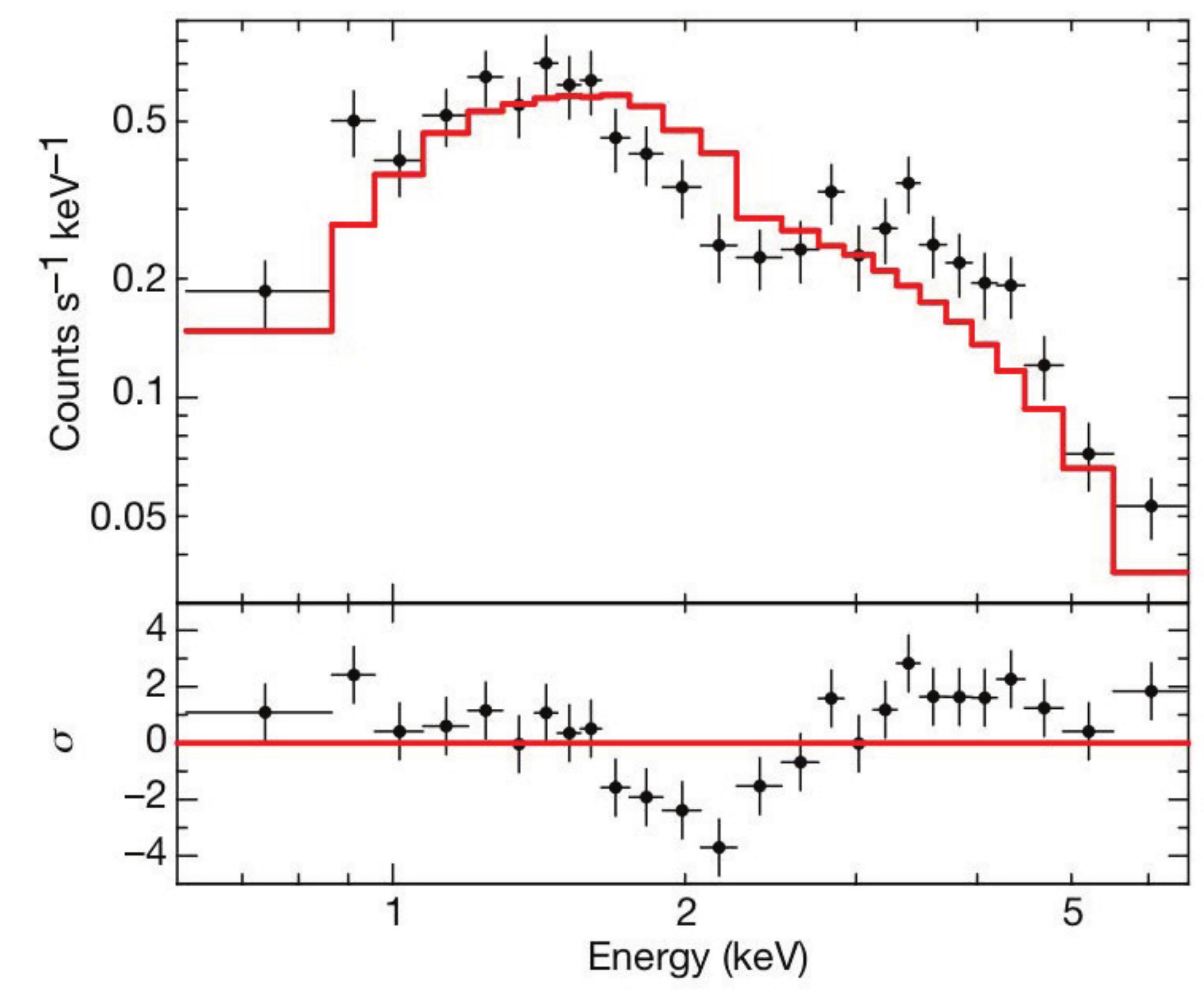}
    \includegraphics[width=0.4\textwidth]{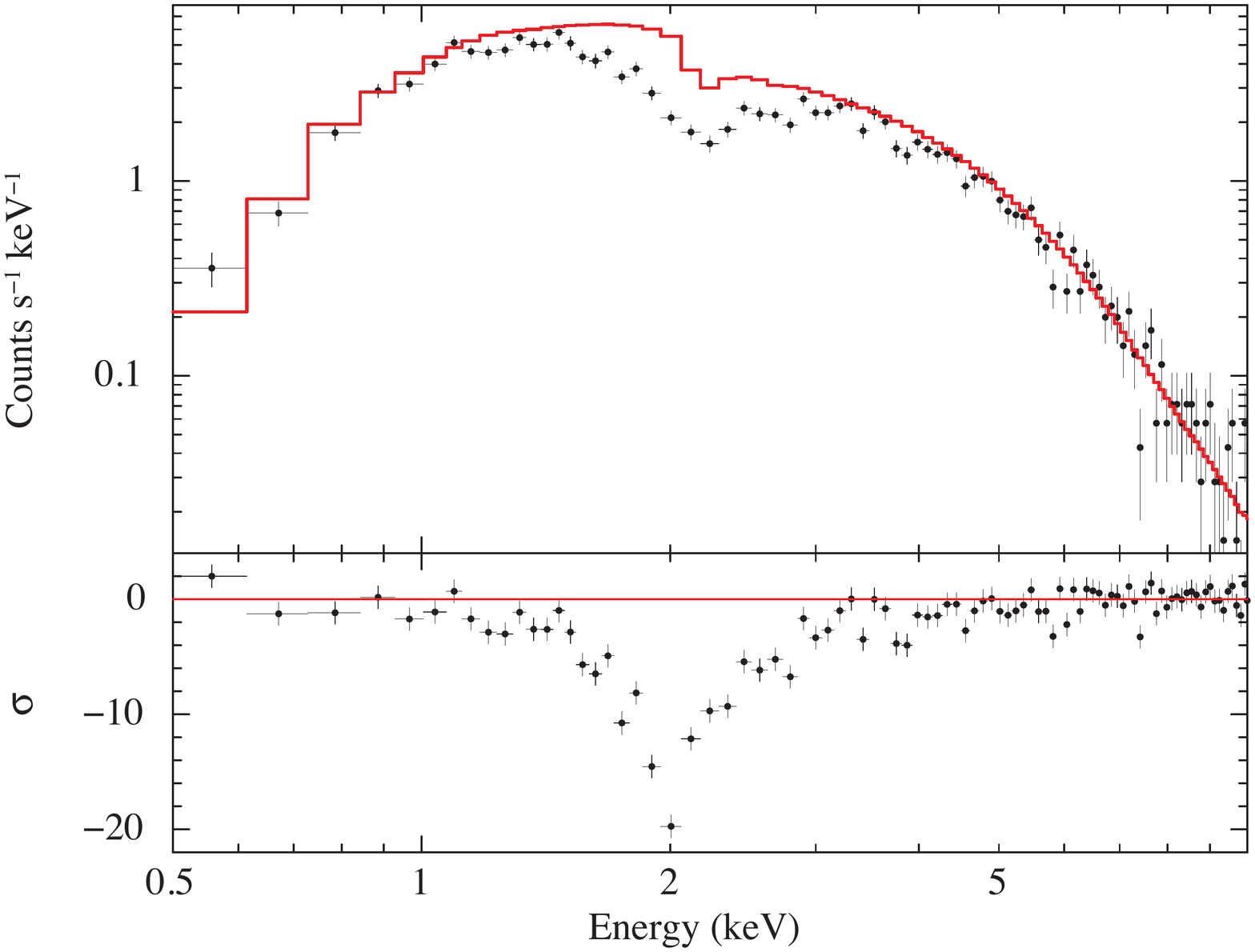} \caption{The top
    left figure shows an artist's impression of the magnetic field topology
    inferred for the magnetar SGR 0418+5729 thanks to the discovery of a
    prominent cyclotron line around 2 keV~\citep{ref42}. The top right figure
    shows the cyclotron feature variation as a function of the magnetar
    spin-phase in a pulse-phase resolved spectrum as observed by \xmm.
    The EPIC-pn spectrum corresponding to a specific phase interval (exposure
    time 613 s) is shown on the bottom left figure. The bottom right figure
    shows the simulated spectrum of the cyclotron feature observed in a
    specific 0.02-wide spin-phase interval and as would be detected in a 613-s
    long observation with \extp SFA. It is evident that \extp would have 
detected
    the same feature at a much larger significance in the same exposure time.
    In both cases, the cyclotron feature is not included in the model used to
    fit the data in order to highlight its presence in the residuals from the
    fits.}
\label{fig:riga1}
\end{figure*}

As a representative simulation, we show in Figure~\ref{fig:riga1} a phase
resolved \extp observation of the cyclotron spectral feature detected in
SGR~0418+5729~\citep{ref42}. This simulation shows that the SFA onboard \extp
will be able to detect, with high significance, the spectral feature as a
function of the neutron star spin. \extp will measure this spectral feature
at the same significance as \xmm in less than 20\% of the exposure time needed
by \xmm, which is particularly important for pulse-phase resolved analysis.

Of the known magnetars, the ideal targets for \extp searches for
spectral lines are the magnetars that are expected to have strong magnetic loops close
to the surface: the youngest objects,  and the transient magnetars during
the outburst peaks.

\subsection{Magnetars in Binaries}
\label{binary}

Magnetars have been so far
associated only with isolated sources. However, accreting magnetars in binary
systems may exist. Based on our knowledge of the 
phenomenology of isolated magnetars, it has been proposed to search for accreting magnetars by
looking
for three signatures~\cite{ref45,ref46}: (1) magnetar-like bursts in an
accreting neutron star system; (2) a hard X-ray tail above $100\,\rm keV$; (3)
an ultra-luminous X-ray pulsar (like that observed in M82,~\cite{ref47}). In
fact, accreting magnetar systems may form a significant fraction of
ultra-luminous X-ray sources \citep{ref46, ref47a, ref47b}.

With its large area, broad band and high sensitivity resolution, \extp would be
able to detect magnetars in binary systems in the future. Repeated
magnetar-like bursts in an accreting system would be strong evidence for
accreting magnetars and magnetism-powered activities~\citep{ref48}. 
Detecting such activity on top of the accretion powered flux requires,
however, large effective area and long exposure times - provided for the first
time by \extp. The transition between the accretion phase and the propeller
phase might also be used to uncover magnetars in binaries~\citep{ref45,ref49},
and detection of such a transition on short timescales also requires large effective area.
Finally, the spin period evolution of X-ray pulsars monitored by \wfm
might also be used to find strongly magnetized objects.

\section{Strong Magnetic Fields at work: accretion and rotation powered pulsars}
\label{AccrP}
\subsection{Introduction}
Emission from X-ray pulsars is powered by accretion of ionized gas onto 
strongly magnetized (B$\sim$10$^{12-13}$\,G) rotating NSs. In this 
``magnetospheric'' type of accretion, the plasma flow is disrupted by the 
rotating magnetosphere far away from the neutron star at the 
magnetospheric radius, $r_{\rm m}\sim10^9$\,cm, and funnelled along the 
field lines to the polar areas of the NS \cite{ref50,ref51}. Upon impact 
with the surface, the kinetic energy of the accretion flow is converted to 
X-ray emission, which appears to be pulsed due to the rotation of the NS.

The observational appearance of X-ray pulsars is diverse. It is defined by 
several factors including the nature and evolutionary status of the donor star,
the parameters of the binary system, and of the neutron star
\cite{ref52,ref51,ref53}. In particular, the magnetic field strength and
configuration define the geometry and physical conditions within the emission
region and thus strongly affect the observed X-ray spectrum. On the other hand,
the interaction of the accretion flow with the magnetosphere mediates the accretion
rate, thus affecting detectability and variability properties of the 
pulsar, and this also defines the spin evolution of the neutron star.

A key probe of the magnetic field intensity is provided by the electron 
CRSFs observed in spectra of many X-ray pulsars. These line-like 
absorption features are associated with the resonant
scattering of X-rays by plasma electrons in their Landau levels:
$E^{(n)}\simeq(n+1){e \hbar B \over{m_ec}}$. The electron cyclotron energy
depends only on the magnetic field in the line forming region
$E_{cyc}=11.6 (B/10^{12}~{\rm G})$~keV, so observations of the CRSFs 
provide a unique
direct probe of the magnetic field in the line forming region. Given the 
typical NS
fields of $\sim10^{12}$\,G, the CRSFs are observed at hard X-ray energies in
X-ray pulsars (see \cite{ref54,ref55} and references therein).

The observed energy and shape of the CRSFs are, however, also affected by the
geometrical configuration of the emission region and the orientation with
respect to the observer. Indeed, given the strong angular dependence of the
scattering cross-sections, observed spectra are expected to be significantly
different when most of the photons are propagating predominantly along as opposed to
predominantly across the magnetic field. For a rotating pulsar, this implies a
strong dependence of the observed X-ray spectra on pulse phase, which can be
used to probe the geometry and physical properties within the emission region.

Note that the geometry of the emission region can change with the accretion
rate, so the spectra of X-ray pulsars also change with luminosity. It was proposed
by \cite{ref56} that a major change in the geometry of the emission region
must occur close to the local Eddington luminosity $L_E$, at the so-called 
critical luminosity $L_c\sim10^{37}$erg\,s$^{-1}$ \citep{ref57,ref58}. At low
luminosities ($L<L_c$) the influence of the radiation on the infalling matter
is negligible, so the kinetic energy of the accretion flow is only released
upon the impact with the NS surface, forming hot compact polar caps. 
The opacity of
strongly magnetized plasma is lower along the field lines, so the 
Comptonized
X-rays escape predominantly upwards along the field, and form a so-called
``pencil-beam'' pattern. At higher luminosities 
($L>L_c$), a
radiation dominated shock rises above the neutron star surface, forming an
extended accretion column \cite{ref56, ref59, ref60}. In this case, the 
photons
can only escape through the sidewalls of the column, and a ``fan'' emission
pattern is predicted.

As already mentioned, the pulse phase dependence of X-ray spectra is expected 
to be drastically different for the two geometries, so pulse-phase 
resolved spectroscopy is commonly used to probe the accretion regime. 
Unfortunately, interpretation of the results is complicated by the fact 
that both poles of the neutron star might be visible at all pulse phases, 
and it is not trivial to disentangle their respective contribution. 
Analysis of pulse-phase resolved spectra in a range of luminosities can 
be used to reconstruct the intrinsic spectra and beam patterns of the two 
poles \citep{ref67k}, which is an essential step in interpretation of the 
observed spectra. Such analysis requires, however, several additional 
assumptions and high quality observations, particularly below the critical 
luminosity. This implies low observed fluxes, and is thus challenging 
for existing facilities.

Scattering of the emerging X-rays in strongly magnetized plasma 
is also expected to induce a high degree of linear polarization of the 
observed emission. Polarisation properties are expected to  
depend strongly on the orientation with respect to the magnetic field, and to be sensitive 
to the emission region geometry. Polarimetric observations are thus highly 
complementary to the more traditional techniques, and can be used to 
reduce the number of assumptions required to reconstruct intrinsic beam 
patterns and spectra of X-ray pulsars. \extp will be the first mission to 
provide simultaneously both X-ray polarimetry and large effective area 
essential for pulse-phase and luminosity resolved spectroscopy.

\subsection{Pulse phase and luminosity resolved spectroscopy}
In high luminosity sources most of the observed emission is expected
to come from the accretion column with a ``fan'' beam pattern.  The observed CRSF
energy in this case depends on the height of the visible part of the accretion
column. The magnetic field is weaker away from the NS surface, so the
energy of the CRSF formed in the upper parts of the column is lower. Variations of
the observed line energy with pulse phase are therefore believed to 
be associated
with changes in visibility of different parts of the column, partially 
eclipsed by
the neutron star.

The height of the column increases with luminosity, so a luminosity dependence of
the CRSF energy is also expected. In particular, an \textit{anti-correlation} of
the observed line energy with flux is most often attributed to the increase of
the accretion column height with luminosity. Alternatively, the line may form
in the NS's atmosphere illuminated by the accretion column. In this case 
the
observed anti-correlation of CRSF energy with flux in super-critical sources
can be attributed to a change of the illumination pattern associated with
column height variations \cite{ref61,ref62}. In either case, this effect 
can be
probed through analysis of the dependence of the observed line energy on
luminosity, and used to constrain the physical properties of the accretion
column.

On the other hand, a \textit{correlation} of the CRSF energy with flux has been
reported for low luminosity sources \cite{ref66}.  Here the observed
change of the CRSF energy is also believed to be due to a change of the line forming
region height, which is expected in this case to decrease with luminosity
\cite{ref66}. Alternatively, the observed correlation of the line energy with
flux in low luminosity sources can be attributed to Doppler shifts due to
variation of the accretion flow velocity \cite{ref58a}.

\begin{figure*}[ht!]
    \centering
    \includegraphics[width=0.32\textwidth]{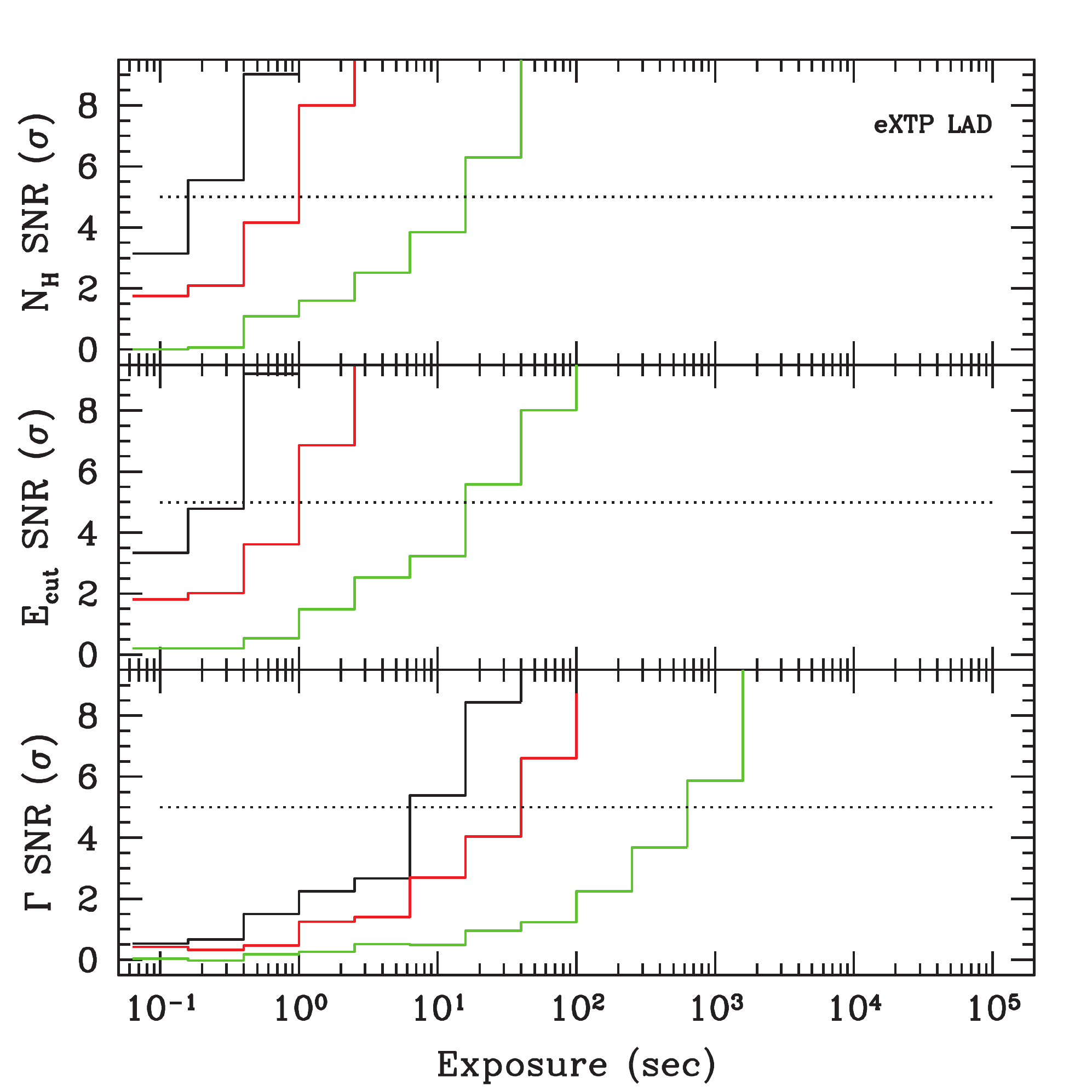} \includegraphics[width=0.32\textwidth]{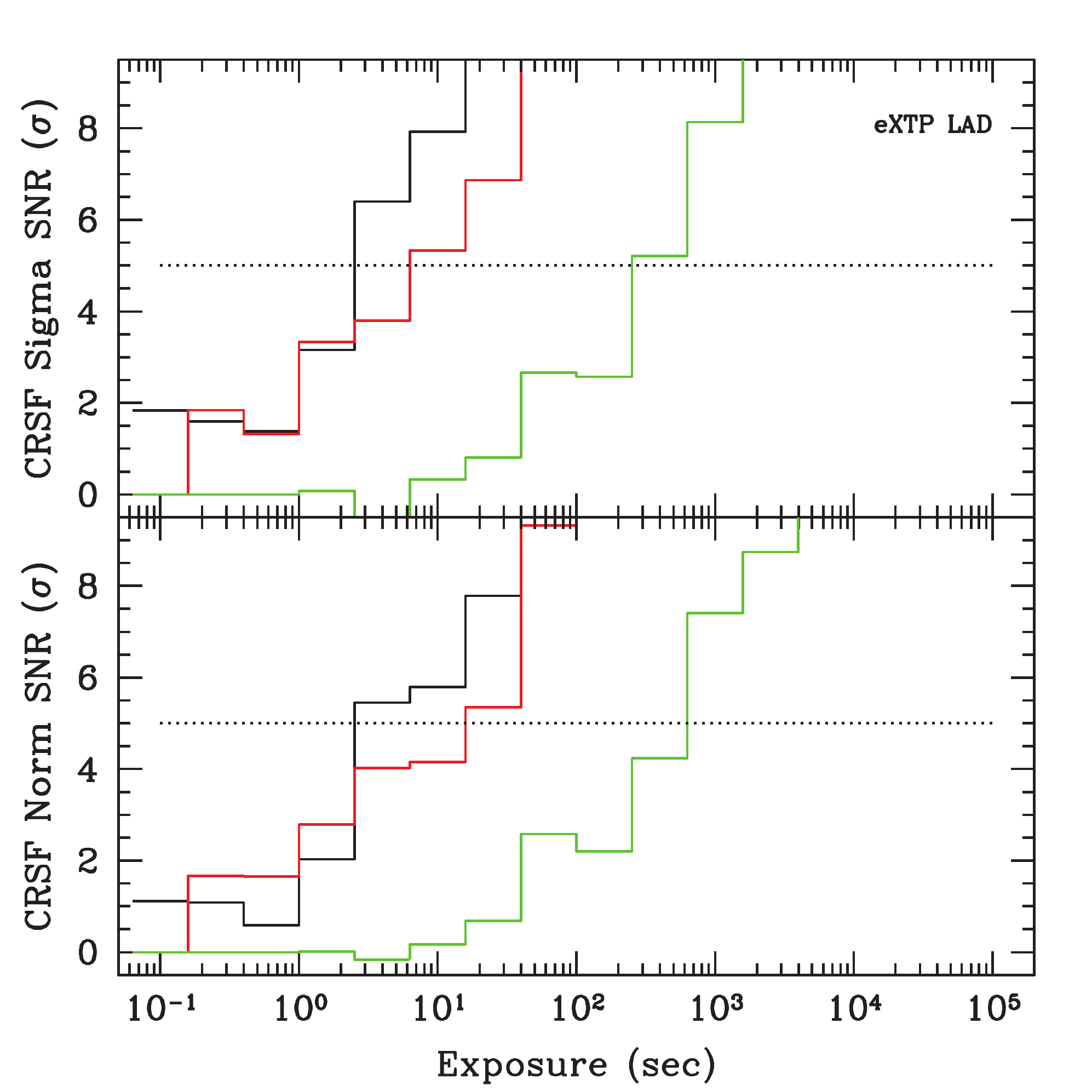}
    \includegraphics[width=0.32\textwidth]{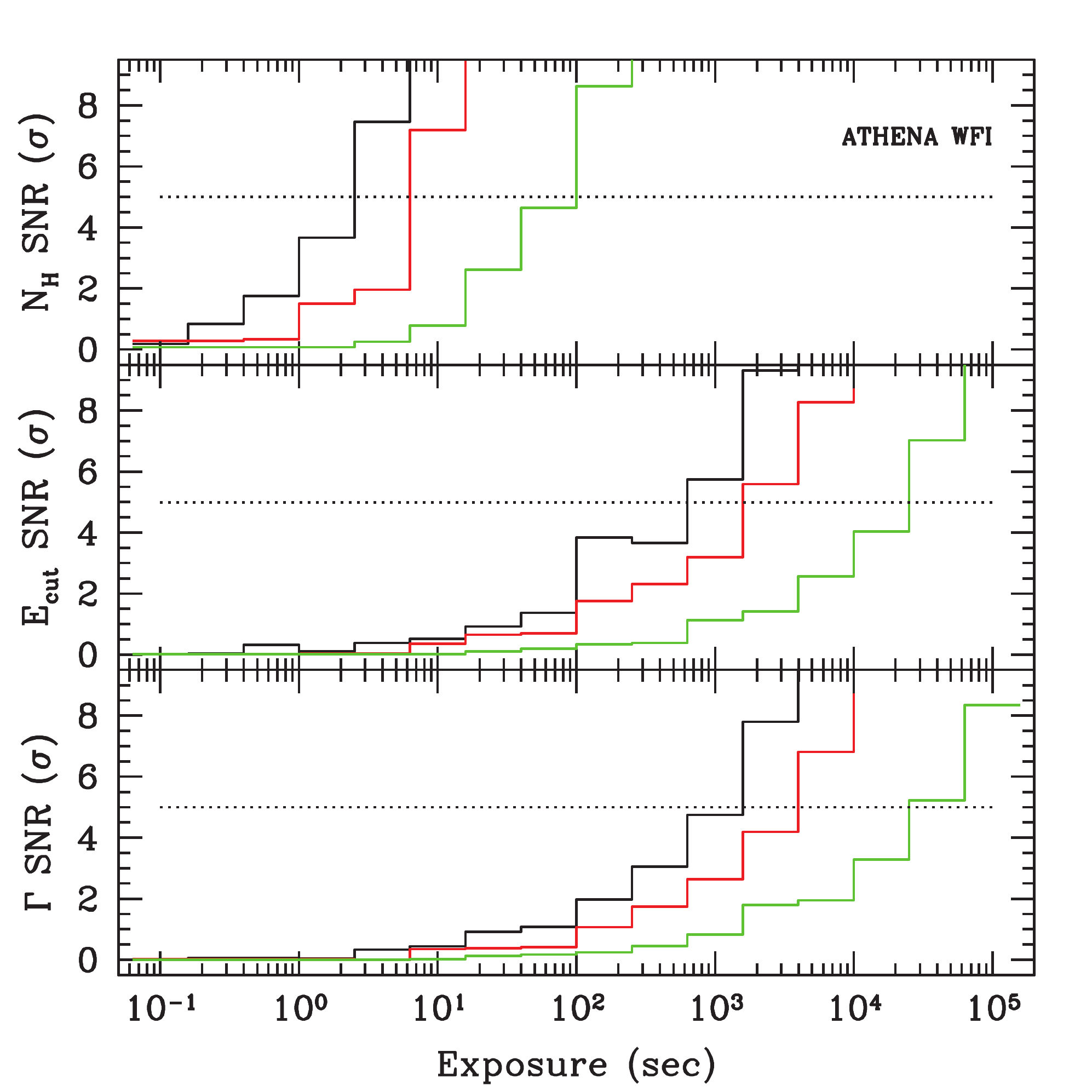}
    \caption[]{Left panel: an example of the \extp \lad capability to 
reconstruct the
    spectral shape of an accreting X-ray pulsar in very short integration
    times. Simulations were performed using the results of a BeppoSAX
    observation of Vela X-1 \cite{ref71}. Each curve gives the significance of
    a spectral parameter (defined as the ratio between the best fit parameter
    and its uncertainty) as a function of the exposure time. The curves are
    color coded according to the assumed flux (black: 500~mCrab, red:
    150~mCrab, green: 15~mCrab). 
    Middle panel: Same as before but showing the \lad
    capabilities in the case of 4U~0115+634. For this source, we show the
    significance of the CRSF detection at 12.8~keV using the spectral model
    proposed by \cite{ref55}. 
     ``Norm'' is the CRSF
    normalization and ``Sigma'' its width. We do not plot the significance of
    the CRSF centroid energy, as it is always determined at very high
    significance. The \lad can indeed measure the CRSF centroid energy at a
    significance level of 100, 20, and 3$\sigma$ for fluxes of 500, 150, and 15
    mCrab, respectively in only 0.1~s. Right panel: Same as the left panel, but
    for Athena/WFI. This shows that similar studies are only possible if 
the \lad
    large collecting area is available. Other currently planned X-ray
    instruments would need overly long exposure times to achieve accurate spectral
    measurements.
    }
\label{LAD-SNR}
\end{figure*} 

The large effective area and superb energy resolution of \extp \lad will
allow us to monitor luminosity-related changes of the observed CRSF energy 
in
transient sources, with unprecedented accuracy.  As illustrated in
Fig.~12 or the \extp observatory science white paper, \extp is capable of constraining line parameters with high
accuracy across a wide range of luminosities, which will enable detection 
of 
the accretion regime transition in many sources sources.

In addition, \extp will bring pulse phase
and luminosity resolved spectroscopy to an entirely new level.v Pulse phase 
resolved analysis has so far been hampered by the fact that  with existing facilities, one must
average the source spectrum at a given phase over many individual pulses. This provides, however, information only on
the averaged properties of the emission region, whereas the relevant 
timescale
for intrinsic changes is comparable to or shorter than the spin period of 
the 
neutron
star. Lack of information on spectral variations on short timescales thus
limits the development of theoretical models \citep[see, e.g.,][for a recent
review]{ref65}, as a result of which these topics remain highly debated
\citep{ref66,ref67,ref68,ref57,ref62}.

The large throughput of the \lad will allow us,  
for the first time, to accurately
constrain spectral properties of accreting X-ray pulsars, by using 
integration times much shorter than a single NS pulse. This is illustrated in
Fig.~\ref{LAD-SNR} for two X-ray pulsars prototypes: Vela\,X-1 and 4U~0115+634.
It is worth mentioning that given the optimal response and effective area of
the \lad at 6-8~keV, pulse phase resolved spectroscopy will not be limited to
CRSFs and broadband continuum, but will naturally encompass other spectral features,
e.g., the fluorescence iron-K emission line (simulations show that the typical
iron line observed in Vela X-1 can be constrained by the \lad with an
integration time $\lesssim$10~sec).

The \lad will also be able to perform pulse phase spectroscopy at low
luminosities, a regime not yet investigated due to the lack of
simultaneous broad-band energy coverage, energy resolution, and collecting area
of past and present X-ray facilities. Particularly intriguing is the variation of
the cyclotron line energy with luminosity observed in several sources, see e.g.,
\cite{ref69}.  As already mentioned, such a variation most probably reflects a vertical
displacement of the emitting region in the inhomogeneous magnetic field of the
NS, and has to date been investigated mostly at luminosities
$\gtrsim$10$^{37}$\,erg~s$^{-1}$ with relatively long pointed RXTE and INTEGRAL
observations \citep[see, e.g.,][and references therein]{ref70}. At luminosities
below $\sim$10$^{37}$\,erg~s$^{-1}$ the accretion column is expected to
disappear, so a different dependence of the cyclotron line energy $E_{\rm cyc}$
on flux \cite{ref57} might be anticipated. Such a transition has indeed been
recently observed in a bright transient V~0332+53 \citep{ref57a} with
\textit{NuSTAR}. However, observations with \extp/\lad will allow us not 
only to obtain similar results for a larger number of sources, but also 
to investigate for
the first time the pulse-phase dependence of the spectrum at low luminosities.
Simulations in Fig.~\ref{LAD-SNR} show that the \lad will be able to measure
spectral parameters of the CRSF of accreting pulsars at high significance
($\gtrsim 5\sigma$) in less than several hundreds of seconds, even at
luminosities as low as a few tens of mCrab.

\subsection{Polarisation studies of accreting pulsars}
Photon scattering in magnetized plasma is also expected to lead to
strong polarization of the emerging X-rays. In strong magnetic fields the
radiation propagates in two, linearly polarized modes: the ordinary O- and 
the
extraordinary X-mode, which are parallel and perpendicular, respectively,  
to the B-field. The difference in opacities for the ordinary and 
extraordinary modes implies 
strong birefringence for propagating photons and thus strong polarization 
of the emerging radiation. In their seminal paper, \cite{ref63} showed 
that the X-ray linear polarization depends strongly on the
geometry of the emission region, and varies with energy and pulse phase, 
reaching very high degrees, up to 70\%, for favorable orientations \cite{ref63}.
The accreting pulsars are also among the 
brightest X-ray sources in the sky and are thus a prime target for X-ray 
polarimetric observations.

For photons propagating along the field (``pencil beam'') the weakest
polarization is expected when the flux amplitude is at its maximum. On the
other hand, for ``fan beam'' the amplitude of polarization is maximum when the
flux amplitude is maximum. The phase swing (from positive to negative and
viceversa) of the polarization angle with respect to the flux maxima and minima
also clearly depends on the beaming geometry (``fan'' vs. ``pencil'').  
Phase resolved measurements of the linear polarization thus constitute a
decisive test to distinguish between the two scenarios. In other words,
phase-resolved polarimetry can help to constrain the radiation pattern model
and also the value of the viewing geometry (e.g. the angle that the observer
line-of-sight makes with the pulsar magnetic axis) as functions of the
rotational phase. In addition, since the components forming the overall beam
pattern may dominate at different energies, the energy dependence of the
degree and angle of polarization can be observed. This information can be used
to disentangle unambiguously the contributions of the two NS poles and thus
contribute solving the long-standing problem of radiation formation in
accreting pulsars. For instance, polarization measurements will immediately
test the reflection model for the CRSF formation suggested by \cite{ref62}. We
also observe that the phase swing of polarization angle can be used to infer
the angle between the rotation and the magnetic dipole axes, a free parameter
of the state of the art models for pulse profile formation.

\begin{figure*}[ht]
    \centering
    \includegraphics[width=1.0\textwidth]{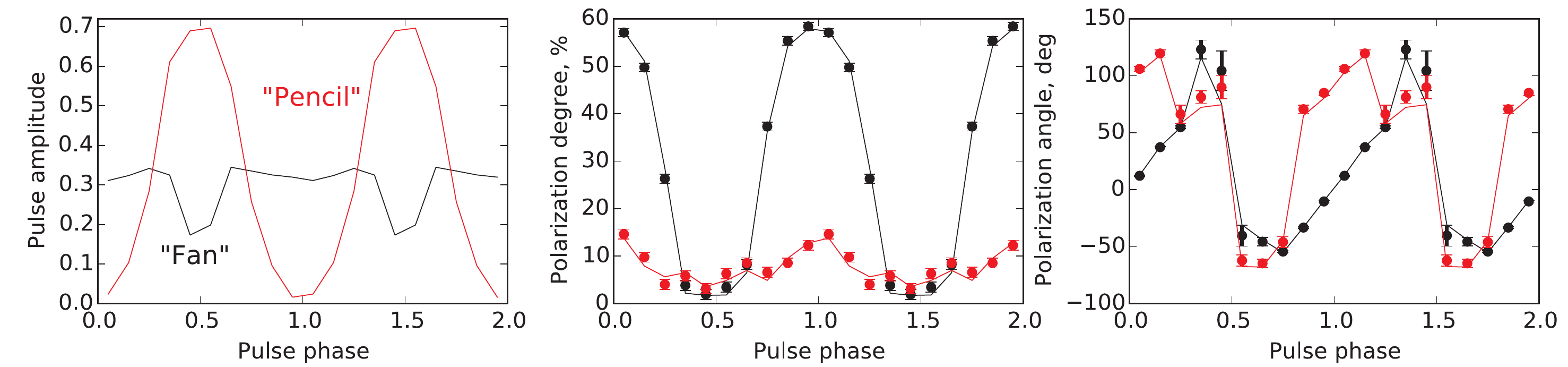}
    \caption{The pulse amplitude, the linear polarization fraction and the
    degree of polarization as a function of the pulse phase are shown, 
assuming
    the spectrum of Vela X-1. An exposure of 7\,ks/phase has been chosen
    together with 10 phases. Total exposure is 70\,ks. A combination of
    $i_1 =45 \circ$ and $i_2=45 \circ$ has also been chosen.}
    \label{fig:DoroshenkoVela}
\end{figure*}

Assuming the simple scenarios predicted by \cite{ref63}, we show in
Fig.~\ref{fig:DoroshenkoVela} a simulation for the PFA onboard \extp
of the bright accreting pulsar Vela~X-1, for the two cases of pencil and fan
beam. Phase resolved linear polarization degree and angle are shown for ten
phase bins with 7\,ks/phase bin exposure, i.e. for a modest total exposure of
70\,ks. In the specific example shown here, it is $i_1 =45^\circ$ and $i_2
=45^\circ$, where the angle $i_1$ refers to the angle between the spin axis and
the line of sight, while $i_2$ corresponds to the angle beween the spin and the
magnetic field axes (see \cite{ref63} for more detail on 
the assumed emission region geometry). The two cases will be neatly distinguished with
the \extp PFA. As expected, the polarization fraction is maximum at the pulse
peak for the fan beam pattern, and minimal for a pencil beam. The skewed
dependence of the polarization angle in the two cases is due to the assumed
geometry (i.e. magnetic/spin inclination), which it can be used to constrain.
The beam pattern emerging from the NS is certainly more complex, and will most
likely be a combination of simpler beam geometries. This will lead to a
more complex phase-dependence of the fraction and angle of polarization.
Moreover, since different beam patterns are expected to be dominant at
different energies, we may observe phase resolved profiles of linear
polarization fraction and angle which changes with energy. The measurements of
the viewing angles $i_1$ and $i_2$, which will allow us to constrain the
geometry of the system, requires a more careful modelling of the emission
region. Several modelling efforts aiming to assess accurately the expected
phase and energy dependence of the polarization properties and flux have been
already started.

X-ray pulsars are among the brightest sources in the sky and are expected to
be strongly polarized, so the characterization of the polarization 
properties 
of
most persistent sources and transients in outburst is within reach of \extp
measurements. This includes both comparatively dim nearby sources and bright
transients throughout the Milky Way. In the case of bright persistent sources,
exposures of the order of $\sim10$~ks will be sufficient for the detailed
studies, while significant polarization will be detected in weak sources 
within
$\sim100$~ks, which still allows phase resolved studies on the Ms 
timescale. In
fact, for several objects the PFA will be sensitive enough to study 
polarization
 in both sub- and super-critical accretion regimes as illustrated in 
Fig.~12 of the \extp observatory science paper.
Such observations would be particularly
important, as the transition from super- to sub-critical accretion is 
expected
to completely change the beam pattern and the polarization properties of the
pulsar, and thus they represent an ultimate test for the theoretical 
models.

\subsection{Inside the NS magnetosphere: microsecond variability in 
High Mass X-ray Binaries (HMXBs)}

In wind-accreting NSs, the plasma is thought to penetrate the magnetosphere
through Rayleigh-Taylor instabilities, in the form of accreting
blobs~\cite{ref73,ref74}. These blobs, after a short radial infall, are
channeled by the magnetic field lines onto the NS polar caps, where they
release their kinetic energy as X-ray radiation. A sort of ``granularity'' 
is
thus expected in the observed X-ray emission~\cite{ref75}, and the energy
release of each shot occurs on microsecond timescales~\cite{ref76}. The
transport of the microsecond pulses through the NS magnetosphere will tend to
broaden them, but photons at energies lower than the cyclotron energy, moving
in a direction parallel to that of the magnetic field, would emerge unscattered
because scattering is drastically reduced~\cite{ref77}. Since the typical
cyclotron resonance energies observed in accreting X-ray pulsars are $E\lesssim
5$ keV, it is expected that 
$\mu$s pulses from the surface of NSs should be detectable relatively
undistorted.

\begin{figure}[H]
    \includegraphics[height=6.3cm]{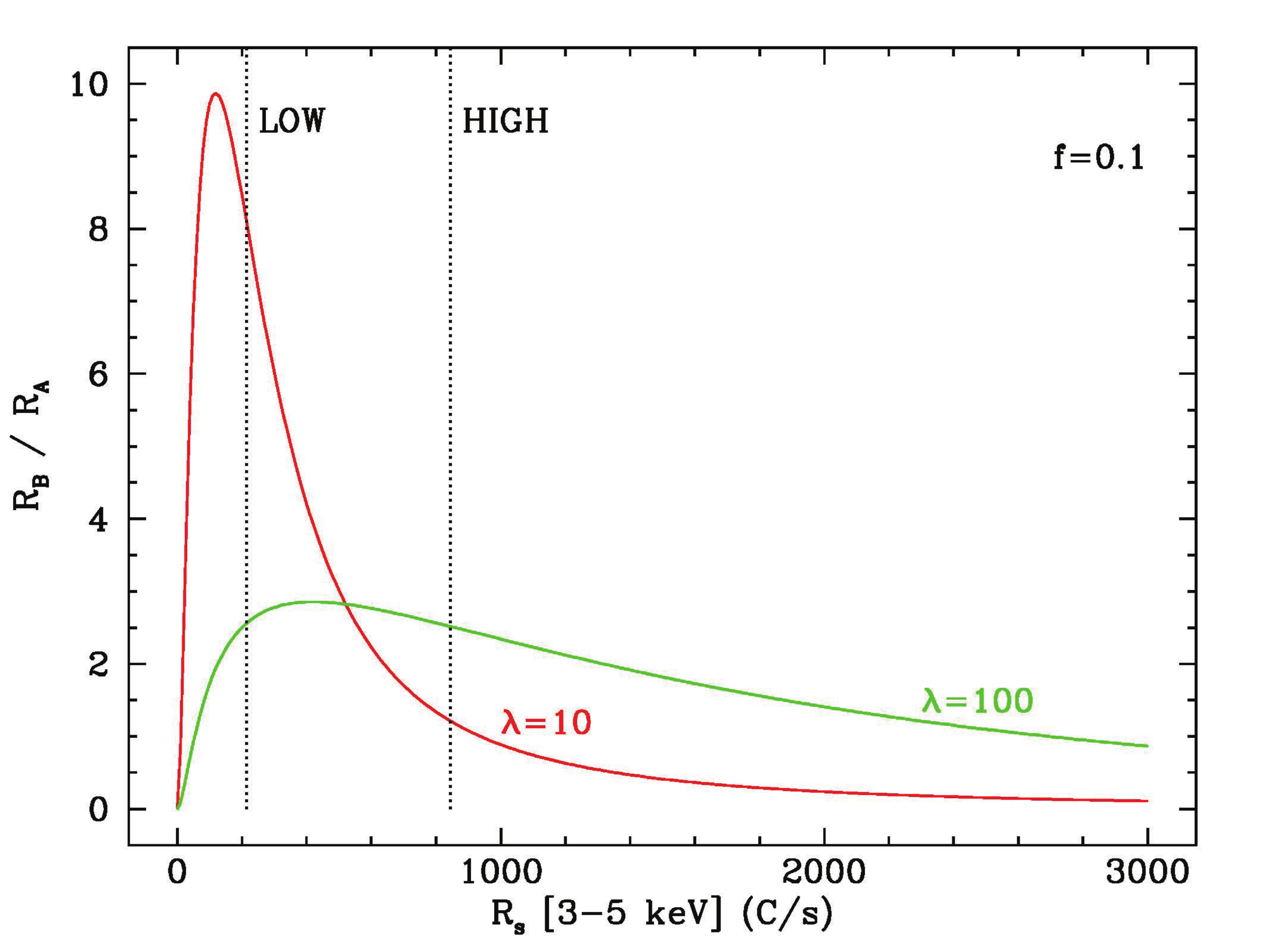} \caption{
    The unique capabilities of \extp in detecting microsecond bursting activity from
    wind-accreting NSs. The ratio between the detection rate $R_B$ of true
    micro-bursts, and the rate $R_A$ of spurious bursts, is plotted as a
    function of the source count rate $R_s$, for the case of a moderate
    granularity (10\%) in the accretion stream. The two vertical lines mark the
    count rates for two Vela\,X-1 luminosity states (low: 15~mCrab, soft
    spectrum; high: 45~mCrab, hard spectrum). $\lambda$ is the rate of
    formation of the blobs at the magnetospheric limit \cite{ref75}. This would
    correspond to the number of rain drops formed in the handrail per unit
    time. $f$ is the granularity component in the accretion flow, here assumed
    to be 10\%.}
\label{dripping-rail} 
\end{figure}

To detect such $\mu$s variability, we plan to use multiple detector 
coincidence.
The \lad is therefore the best suited instrument for this kind of studies, given
its time resolution and the high number of independent detection modules, $N_d$.
The presence of events at $\mu$s time scales can be probed by building a time
interval histogram with 5~$\mu$s resolution. By performing multiple detector
coincidence, it is possible to discriminate between events that occur within
$5\;\mu$s in different detectors. If the X-rays arrive uncorrelated and at a constant
rate, the resulting histogram follows an exponential decay. Any deviation from
that will indicate coherence or correlation among the events. In particular,
microsecond X-ray bursts could be identified as excesses in the time interval
histogram. Following \cite{ref76}, we computed in Fig.~\ref{dripping-rail} the
rate $R_B$ of true bursts from coincidence timing among independent detectors,
and the rate $R_A$ of accidental counts within $5\;\mu$s (e.g., coming from
background fluctuations) for a \lad observation of Vela\,X-1 in the low and high
emission state \cite{ref71}.

According to the results of the simulation, the number of detected events
(bursts plus accidentals) would be significantly greater than that of
accidental events, even for a relatively small granularity of 10\% in both the
low and high emission states. For example, for the low state, the rate of
multiple coincidence detection of true $\mu$s bursts is $R_B$=6..3~c/s, to be
compared to the expected rate of observing accidental coincidences
$R_A$=0.77~c/s \citep[note that this is the rate of true bursts from multiple
detector coincidence, not to be confused with the rate of formation of the
blobs at the magnetospheric limit, $\lambda$, which is in the range 10-100
blobs/s; it is the squashing of these blobs that gives rise to the $\mu$s
variability, as detailed in][]{ref75}. It is important to note that the 
signal to noise ratio (SNR) 
for the detection of microsecond variability scales with the square root of
$\binom{N_d}{2}$. The RXTE PCA had $N_d=5$, therefore the number of pairs of
detection modules with which to perform coincidence was $\binom{N_d}{2} = 10$.
For the \lad, $N_d$=41, and thus the S/N improvement over the PCA is
$\gtrsim$30. No other flown or planned X-ray instrument is thus capable of performing
 such studies efficiently, apart from the \lad.
 
Variability on similar timescales is also expected in HMXBs due to  
so-called ``photon bubble oscillations''. The latter were found to develop below
the radiation dominated shock that terminates the free-fall motion of the
accreting matter, in the accretion column of X-ray pulsars with luminosities
$\gtrsim$10$^{37}$~erg~s$^{-1}$ \cite{ref78}. If convincingly detected, photon
bubble oscillations can provide insights into the structure of the 
accretion
column near the NS surface, and potentially an independent measurement of the
compact object magnetic field \cite{ref78}. Some evidence for the presence of
variability induced by photon bubble oscillations was found in the HMXB Cen
X-3, as two QPOs at 330 and 760~Hz have been
reported in its power spectrum \cite{ref78}. Such features could not be
convincingly confirmed with the 5 PCUs of the RXTE PCA as dead time effects
lowered the counting noise level of the instrument below the value expected
for Poisson statistics, and a counting noise model needed to be included 
in the
fit to the power spectra. In this case, the large number of detectors in the
\lad again provides a unique advantage.

\subsection{Rotation powered pulsars} 
According to the magnetic dipole model, the fast rotation ($\sim1$~ms-$10$~s
periods) of pulsars powers the acceleration of particles in the magnetosphere
and the emission of electromagnetic radiation beamed around the magnetic axis,
producing the characteristic pulsed emission if the magnetic and rotation axis
are misaligned. For this reason they are also referred to as rotation-powered pulsars
(RPPs), to be distinguished from other classes of isolated neutron stars, such
as the magnetars, which are powered by their magnetic energy \cite{ref79}.

Thanks to its features, the \lad on board \extp will make it possible to 
exploit
the diagnostic power of high X-ray time and spectral resolution to 
make progress in
 understanding the radiation emission processes in pulsar magnetospheres.
It is not yet completely established whether the emission at different
energies (gamma-rays, X-rays, optical, radio) originates from different
populations of relativistic particles and different regions (and altitudes) of
the neutron star magnetosphere, nor whether this is related to the occurrence
of breaks in the multi-wavelength power-law (PL) spectra. The light curve
morphologies at different energies are very diverse and encode information about
the orientation of the different emission beams with respect to the neutron
star spin axis. Thus, from their comparison one can map the location of
different emission regions in the neutron star magnetosphere, hence determining
the origin and width of the emission beams at different energies, tracking the
particle energy and density distribution in the magnetosphere, and measuring the
neutron star spin axis/magnetic field orientation. Such a precise
characterisation of pulsar light curves in different energy ranges will be key
to building self-consistent emission models of the neutron star 
magnetosphere
\cite{ref80}.

Over 200 gamma-ray
pulsars\footnote{https://confluence.slac.stanford.edu/display/GLAMCOG/Public+Lis
t+of+LAT-Detected+Gamma-Ray+Pulsars} have been now identified by the Large Area
Telescope (LAT) aboard the Fermi Gamma-ray Space Telescope \citep{ref81,
ref82}. Thus, gamma-ray detections of RPPs outnumber those in any energy range
other than radio, and provide us with a uniquely large and diverse sample to
characterise pulsar light curves at different energies. Focussing, for
instance, on the younger (age lower than $\sim1$~Myr) and more energetic Fermi
pulsars, fewer than 30 of them show pulsed X-ray emission, whereas 
another
40 or so also have (non pulsed) detections in the X-rays \citep{ref83}. 
The \lad on the
\extp mission will accurately measure X-ray light curves and detect pulsations
for Fermi pulsars more efficiently than any other previous mission. We
simulated the sensitivity to the detection of X-ray pulsations at the gamma ray
pulse period with the \lad in the 2-10~keV energy band assuming a single-peak
X-ray light curve with a Lorentzian profile, variable pulsed fraction, and
peak full width half maximum (FWHM). We assumed a power law X-ray spectrum with
a photon index of $\Gamma= 1.5$ and hydrogen column density $N_H$ fixed to the
Galactic value along the line of sight. For instance, in the event of a 
50\% pulsed
fraction and 0.1 FWHM, in 30 ks we can detect X-ray pulsations, compute the
pulsed fraction (10\% error), and the peak phase (0.02 error) for pulsars with
unabsorbed X-ray flux as low as $3\times10^{-13}$~erg~cm$^{-2}$~s$^{-1}$ in
0.5-10~keV. In case of a sinusoidal profile, in 30 ks we can reach the same
accuracy in the light curve characterisation for pulsars with an unabsorbed
X-ray flux as low as $6\times10^{-13}$~erg~cm$^{-2}$~s$^{-1}$. This means that
we can obtain X-ray light curves with the accuracy claimed above for at least
20 of the younger Fermi pulsars currently detected in the X-rays. Moreover,
thanks to the \lad sensitivity up to 50~keV, we can increase by at least a
factor of two the number of pulsars seen to pulsate in the hard X-rays.

Measurements of the X-ray polarization of RPPs, never performed to date but 
possible with the PFA on \extp, will provide additional and
unprecedented information on their highly-magnetized relativistic
environments. For RPPs with X-ray flux as faint as
$5\times10^{-13}$~erg~cm$^{-2}$~s$^{-1}$, we will be able to measure a minimum
detectable X-ray polarization of 10\% (time averaged) with an exposure time of
150~ks, and compare it with the expectations for the X-ray emission mechanisms
(e.g., synchrotron). For the Crab pulsar
($4.4\times10^{-9}$~erg~cm$^{-2}$~s$^{-1}$), we will be able to measure the
time-resolved polarization degree and polarization position angle with an
integration time of 10~ks, and their variation as a function of the pulsar
rotational phase, to be compared with predictions of pulsar magnetosphere
models such as the outer gap, polar cap, and slot gap/caustic gap models,
which predict different swing patterns of the polarization parameters.

Giant Radio Pulses (GRPs) occur in a handful of radio pulsars, mostly
young RPPs. They can be defined as single pulse emission with an intensity that is significantly
higher than the average. Early observations of GRPs indicated a
possible correlation between GRP phenomena and the magnetic field
strength at the light cylinder, whose radius is the distance of the
last closed magnetic field line, possibly pointing to an outer
magnetosphere origin. A correlation between Giant Pulses (GPs) in the
radio and at higher energies has only been observed in the optical for
the Crab pulsar (\cite{Shearer2003}, \cite{Collins2012},\cite{Strader2013}, 
where optical GPs show a $\sim$3\% increase relative to the
average peak flux intensity. 

GPs have not yet been detected in the
high-energy domain, with only upper limits on the relative peak flux
increase obtained at soft ($<$200\%;  \cite{ref88a}) and hard X
rays ($<$80\%;\cite{Hitomi Collaboration2017}), soft
($<$250\%;  \cite{Lundgren1995}, high (\cite{Bilous2011}; $<$400\%)
and very high-energy gamma rays ($<$1000\%; \cite{Aliu2012}). 

From optical observations it appears that the observed energy excess in a GP
is roughly the same in radio as in optical.\ If we scale from
the optical to X-rays we consequently expect to see a similar flux
increase. Our assumption is consistent with the first constraint on a
GP strength in the Crab pulsar in the 15--75\,keV energy range
recently obtained with Suzaku in coincidence with a GRP (\cite{Mikami2014}). 

The high count rate from the Crab pulsar with the \lad and
its $1\,\mu$s absolute timing accuracy will allow studies of individual
pulses. Thus it will be possible to detect, for the first
time, GPs in X-rays from the Crab and in other relatively bright X-ray
pulsars, monitor their correlation with GPs observed simultaneously in
radio with the \textsl{SKA} and determine the correlation
between coherent and incoherent radiation production mechanisms in the
neutron star magnetosphere.  According to our simulations (see \cite{Mignani2015} for details) we should be able to detect a 3\% peak flux increase.

%=====================================QED=============================================

\section{QED}
\label{qed}

\subsection{Introduction}

One of the first predictions of quantum electrodynamics, even before it was
properly formulated, was vacuum birefringence \cite{ref84,ref85}; in
particular that a strong magnetic field would affect the propagation of light
through it. Even when QED was formulated more carefully over the following two
decades \cite{ref86}, this prediction remained robust. In a weak magnetic 
field, the
difference in the index of refraction between photons polarized in the  
O-mode and in the X-mode is simply \cite{ref87} $$ \Delta n = 
\frac{\alpha}{4\pi} \frac{2}{15} \left (
\frac{B}{4.4\times 10^{13}~\mathrm{G}} \right )^2. $$ For terrestrial fields,
this is vanishingly small and has not yet been measured in the eighty years
since the prediction. On the other hand, astrophysically this effect can be
important for neutron stars, black holes and white dwarfs \cite{ref88,ref89}.
For these objects, although the difference in index of refraction is still much
smaller than unity, the combination $\Delta n (l/\lambda)$, where $l$ is the
length over which the index of refraction changes and $\lambda$ is the
wavelength of the light, may be very large. This means that the 
polarization 
states of the light evolve adiabatically, so light originally 
polarized in 
the X-mode will remain in the X-mode even if the direction of the field 
changes. From the point of view of the observer, the direction of the 
polarization will follow the direction of the magnetic field as the light 
propagates through it.

The crucial connection to \extp is that magnetized neutron stars and accreting
magnetized white dwarfs (polars) naturally produce polarized X-ray 
radiation,
and furthermore the field strengths are sufficiently strong and the length
scales sufficiently large that vacuum birefringence decouples the 
polarization
modes, which can have a dramatic effect on the observed polarization. 
Although black holes themselves do not harbor magnetic fields, recent 
calculations also
indicate that the vacuum polarization is important for the propagation of
radiation near black hole accretion disks if magnetic fields provide the bulk
of the viscosity \cite{ref89}. Our first focus will be neutron stars.

\subsection{Neutron Stars}

\begin{figure*}[ht!]
\Centering
    \includegraphics[width=0.8\textwidth]{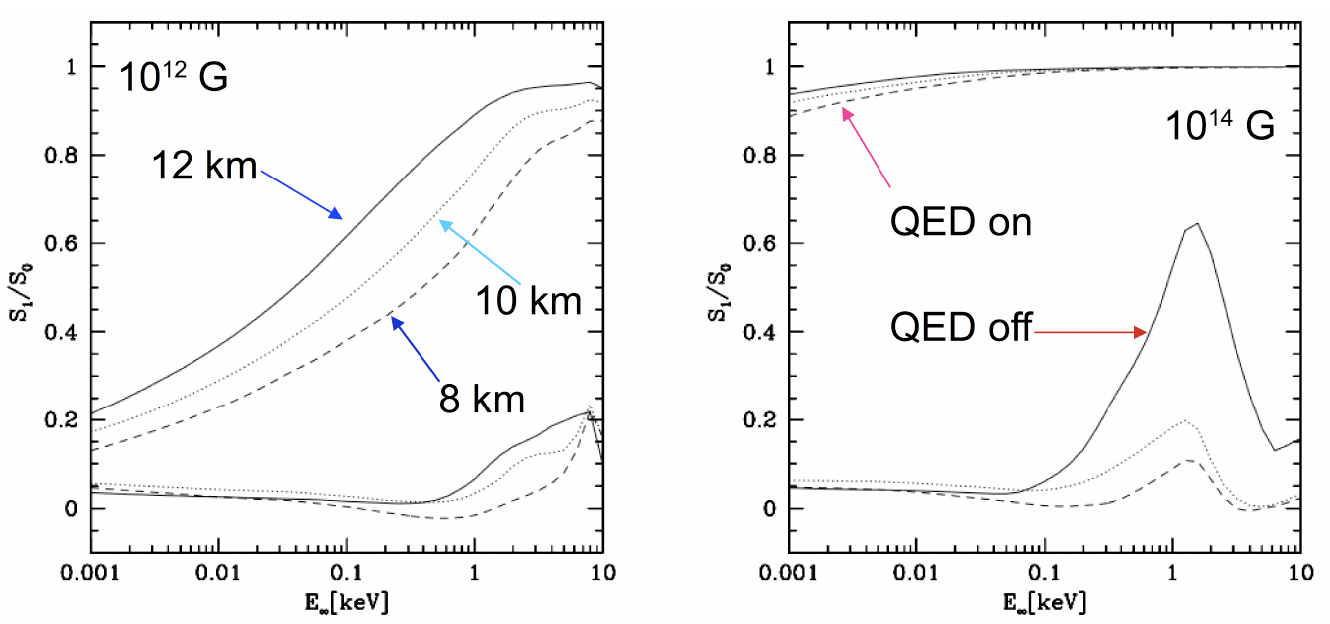}
    \caption{The expected fraction of linear polarization from the surface of
    strongly magnetized neutron stars with hydrogen atmospheres and effective
    temperatures of $10^{6.5}$~K. (from \cite{ref90}.}
\label{fig:surface_magnetar}
\end{figure*}

Vacuum birefringence is strongest for magnetars, whose fields range 
from
$10^{14}$ to $10^{15}$~G. It is for these objects that available calculations
are the most comprehensive. Vacuum birefringence increases the expected linear
polarization of the X-ray radiation from the surface of a neutron star 
from
about 5-10\% to nearly 100\% \citep{ref90}. It is nearly as strong for neutron
stars with more typical magnetic fields of $10^{12}$~G.
Fig.~\ref{fig:surface_magnetar} depicts the results for a typical magnetic
field and for a magnetar. For a magnetar, at the high photon energies observed
by \extp, the effect essentially saturates: without QED, the observed
polarization is small, at most 20\%, and depends on the radius of the 
star,
while with QED the polarization is nearly 100\%. For a more weakly 
magnetized
star, the polarization fraction from the entire surface is also much 
higher than
without QED but, tantalizingly, it is not saturated and depends on the
radius of the neutron star.

Depending on the particular source, the emission from thermal magnetars 
typically peaks around 2-3~keV, where the \extp PFA will have an effective area 
of about 900~cm$^2$, with the gas-pixel detector (GPD) for polarization 
sensitivity. The 100\% polarized radiation from the magnetar will result 
in about 40\% modulation in the detector.

\begin{figure*}[ht!]
    \Centering
    \includegraphics[width=0.4\textwidth]{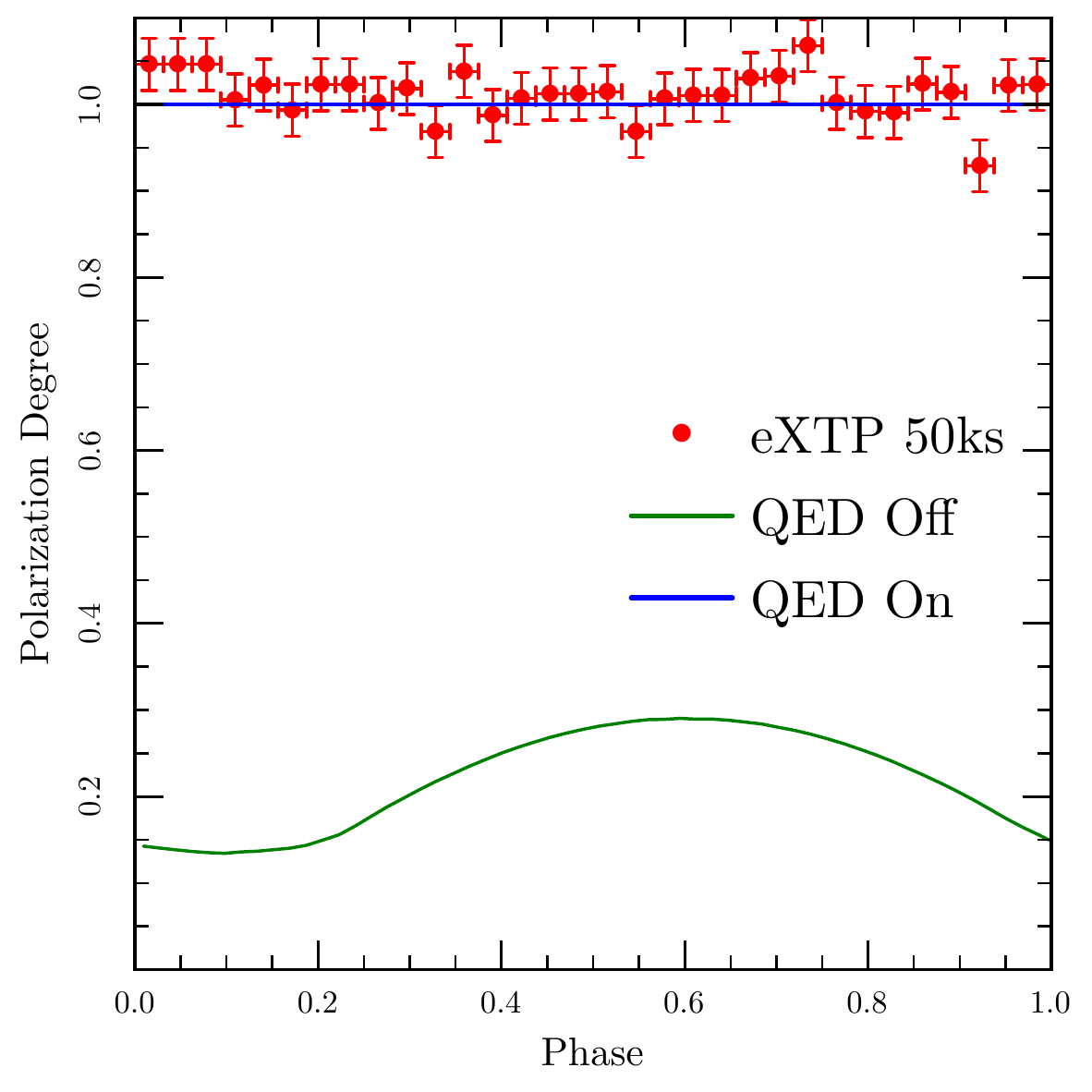}
    \includegraphics[width=0.4\textwidth]{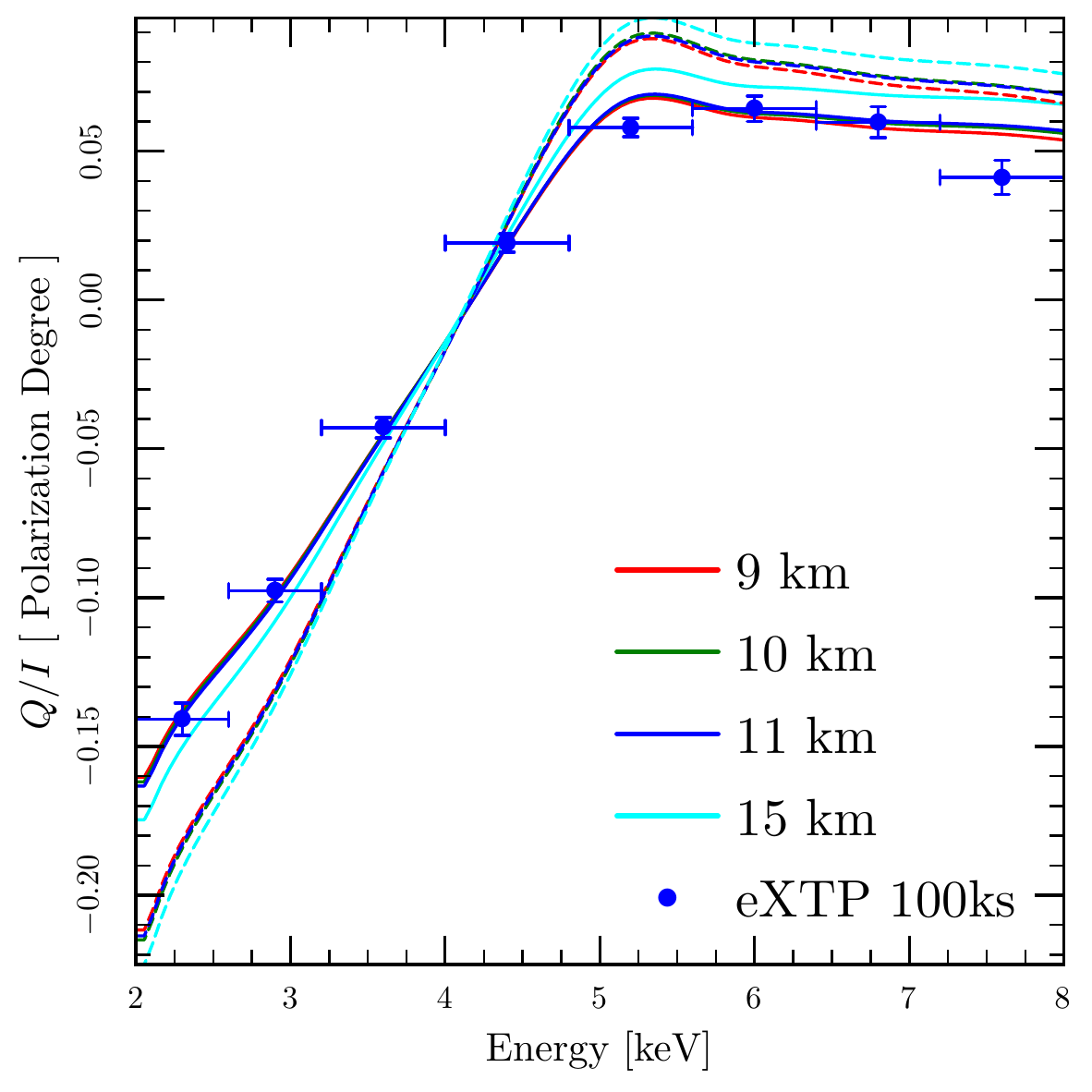}
    \caption{Left: The total degree of polarization as a function of phase
    expected from AXP 4U~0142+614 with and without QED averaged over energy.
    Right: Total polarization averaged over the rotation of Her X$-$1 (see text).
    The \extp simulation assumes a radius of 10~km and QED. The solid curves
    include QED birefringence and the dashed curves do not.}
\label{fig:4u_herx1}
\end{figure*}

Using the AXP 4U~0142+614 as an auspicious example, because its thermal
emission is strong, we would expect to detect about 0.5 photons per second with
energies between 2 and 4~keV with the PFA. The left panel of
Fig.~\ref{fig:4u_herx1} depicts the energy-averaged polarized fraction 
expected
from AXP 4U~0142$+$614 and the results of a 50~ks simulated observation with
\extp, as calculated in a simple model in which the whole NS surface is
emitting at a homogeneous temperature. As it can be seen, this shows that the
QED-off hypothesis can be rejected with a very high significance.

Recently, an optical polarimetry observation of an X-ray dim isolated
neutron star (XDINS), RX J1856.5$−$3754 (with the magnetic field of
$\sim10^{13}$\,G), performed with the VLT, showed a phase-averaged 
polarization
fraction, PF$\sim$16\%, consistent with the first ever evidence for QED vacuum
birefringence induced by a strong magnetic field \cite{ref90a}. This important
result can be confirmed by performing systematic polarimetry observations
in the X-ray band, and the most promising sources are the brighest 
magnetars, either persistent or transient magnetars (hereafter TMs). 

\begin{figure*}[t!]
    \includegraphics[width=0.49\textwidth]{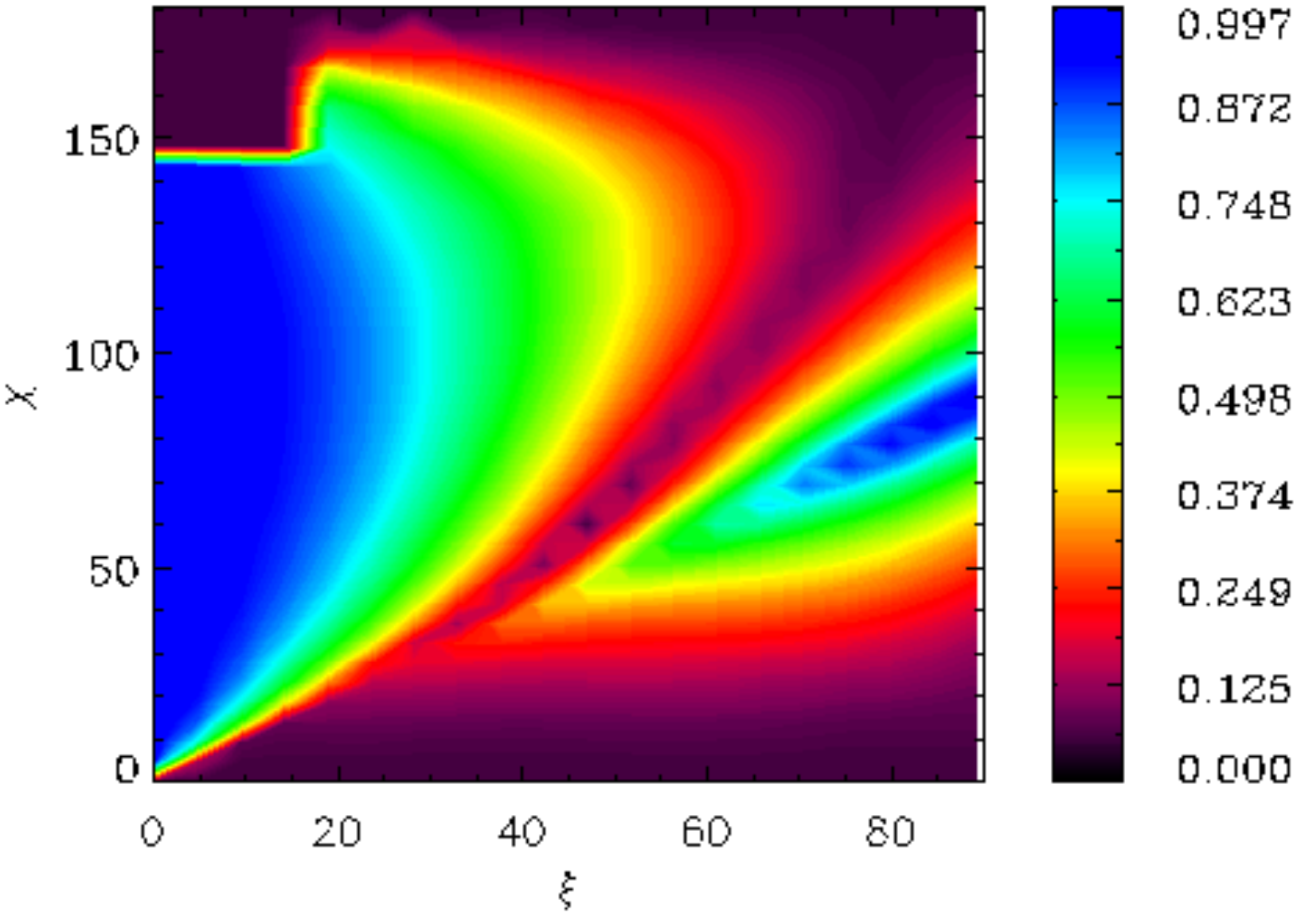}
    \includegraphics[width=0.49\textwidth]{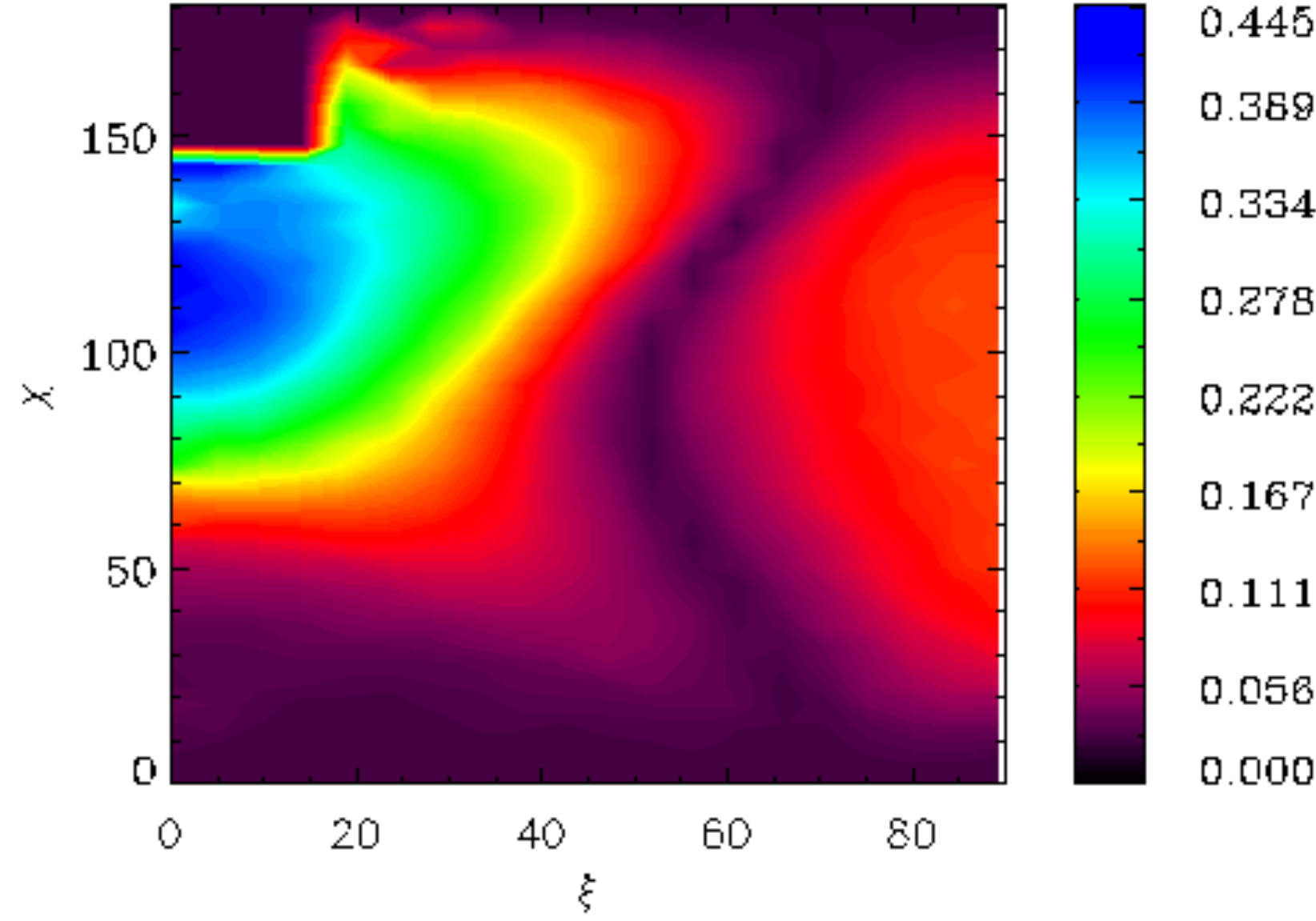}
    \caption{Left: phase-averaged PF for the thermal radiation emitted 
from a TM at the peak
of the outburst (without vacuum birefringence) computed for different
values of the angles $\xi$ (between the magnetic axis and spin axis) and
$\chi$ (between the line of sight and spin axis). The radiation is computed
for a magnetized NS atmosphere, with $B\sim10^{14}$\,G and assuming
emission from one hot polar cap covering $15\%$ of the NS surface.
Right: Same as in the left panel but accounting for the QED effects.}
\label{fig:qed1}
\end{figure*}

\begin{figure}[H]
\includegraphics[width=0.5\textwidth]{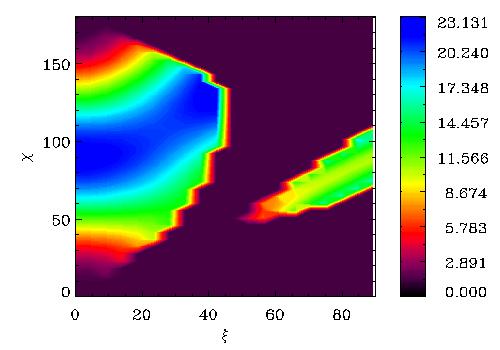}
    \caption{Expected variation of the phase-averaged
    PA for TMs during a magnetospheric untwisting of 0.5\,rad (as expected
    during the outburst decay, for details see \cite{refG17}), in the 2−6 keV
    energy range.}
\label{fig:qed3}
\end{figure}

\begin{figure}[H]
    \includegraphics[width=0.45\textwidth]{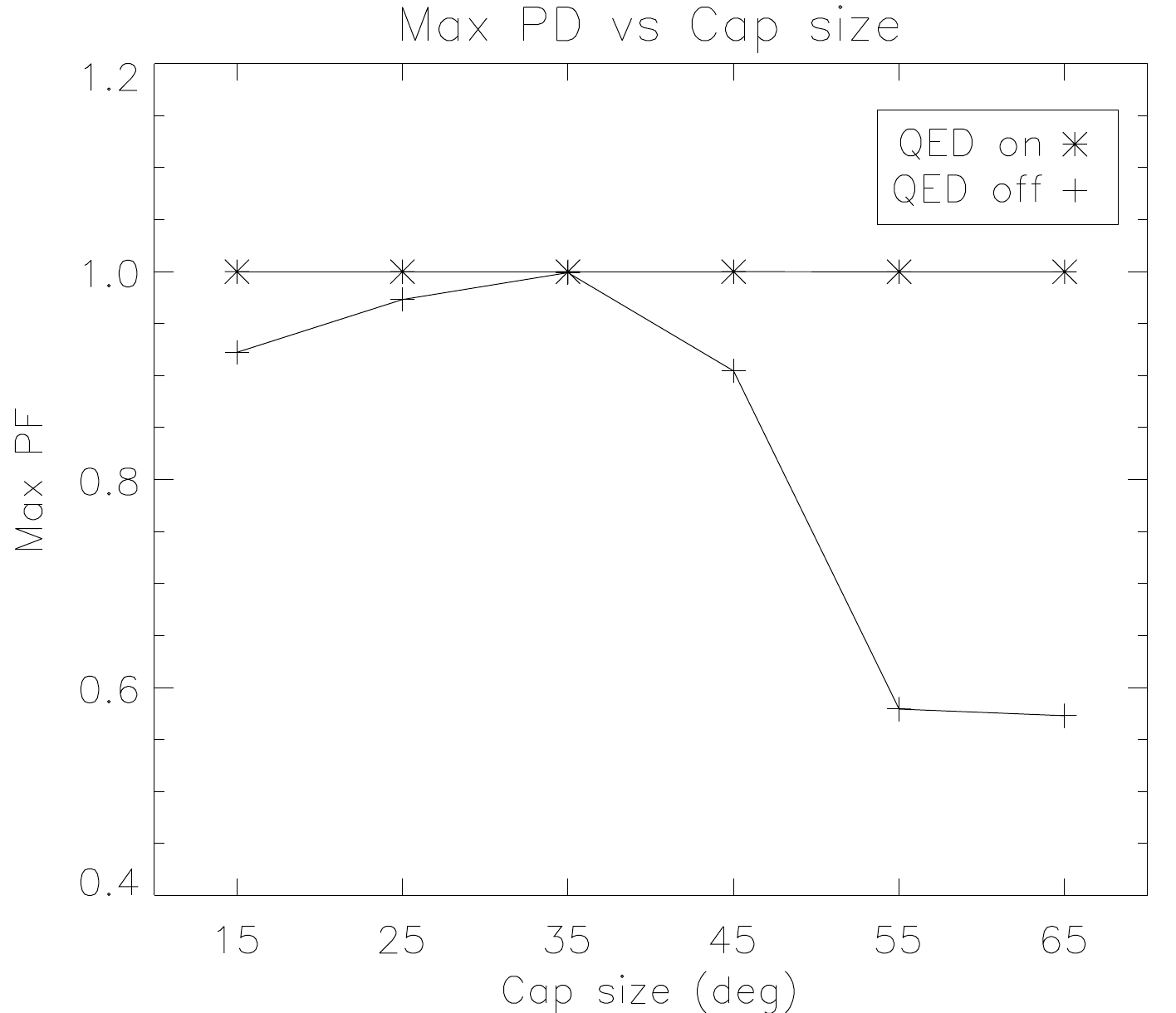}
\caption{Maximum phase-averaged PF for different sizes
of the polar cap (semi-angle of the polar cap). The line with asterisk symbols
(*) corresponds to the case in which vacuum birefringence is operating. The
line with crosses symbols (+) shows the case in which vacuum birefringence is
not present. The radiation at the star surface is computed for a magnetized,
pure-H atmosphere with $B_p=10^{14}$\,G and temperature $T=0.5$\,keV.}
\label{fig:qed4} \end{figure}

We have modeled the 
polarization properties of TMs in order to study the
phase-averaged polarization fraction (PF) and polarization angle (PA) 
during
the TM flux decay, adapting the method used by \cite{ref90ab}. Based on
the observational studies of representative sources such as XTE~J1810$−$197 and
CXOU~J164710.2$−$455216 \cite{ref90abc}, we considered a TM with one hot polar
cap, located at one of the magnetic poles, covering 15\% of the star's surface.
We assume that, at the flux peak, the magnetosphere has a twist angle
$\Delta\Phi=1.0$\,rad and the hot polar cap has a uniform temperature of
$T=1$\,keV. When quiescence is reached, the magnetosphere has untwisted by
$\Delta\Phi=0.5$\,rad and the polar cap cooled down to $T=0.5$\,keV. These
values are reminiscent of those inferred by \cite{ref90abc}, when analyzing
the outburst decay of XTE~J1810$−$197 and CXOU~J164710.2−455216. Once again,
simulations (see \cite{refG17,refG17b} for all details) show that the
super-strong magnetic field surrounding TM can boost the observed PF of the
thermal radiation, via the effect of vacuum birefringence, up to $\sim99$\%
(considering the most favourable viewing geometry $\chi\sim90^{\circ}$ and
$\xi\sim0^{\circ}$, see Fig.~\ref{fig:qed1}). 
We also found that when vacuum birefringence is operating, the value of 
PA can change by up to 23 degrees during a magnetospheric untwisting of 
$\sim$0.5\,rad (as typically expected from outburst onset until 
quiescence, Fig.~\ref{fig:qed3}), offering an interesting test for the 
magnetar model. If vacuum birefringence were not present and the polar cap 
shrinks during the TM flux decay, we found that the PF would increase from 
∼ 65\% up to ∼ 99\% (Fig.~\ref{fig:qed4}). If instead vacuum 
birefringence is operating, the PF stays almost constant, independent of 
the cap size. This is an interesting result, since the detection of a 
nearly constant PF as the decay proceeds will provide further evidence for 
the presence of vacuum birefringence.

\begin{figure*}[ht!]
    \includegraphics[width=0.5\textwidth]{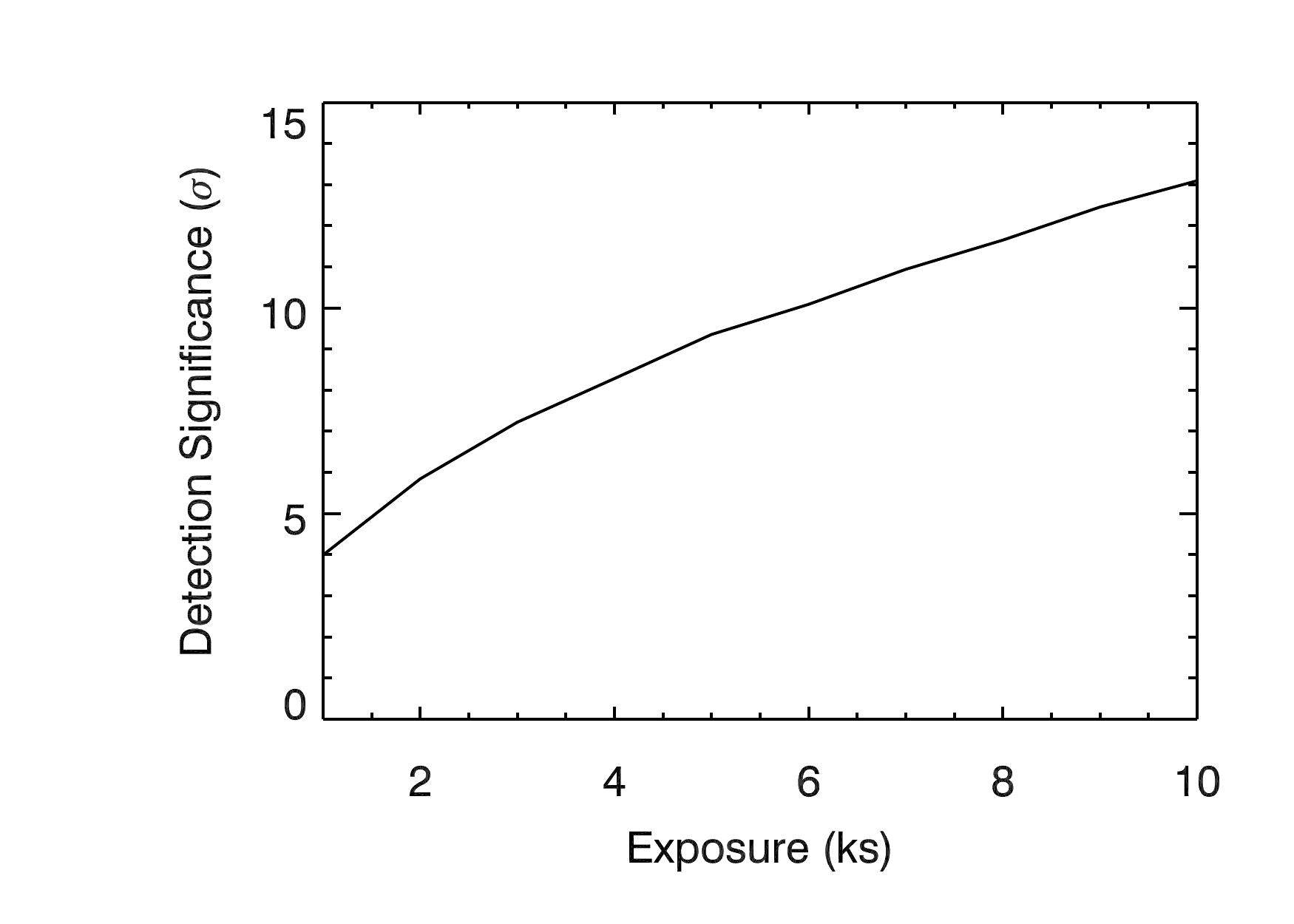}
    \includegraphics[width=0.5\textwidth]{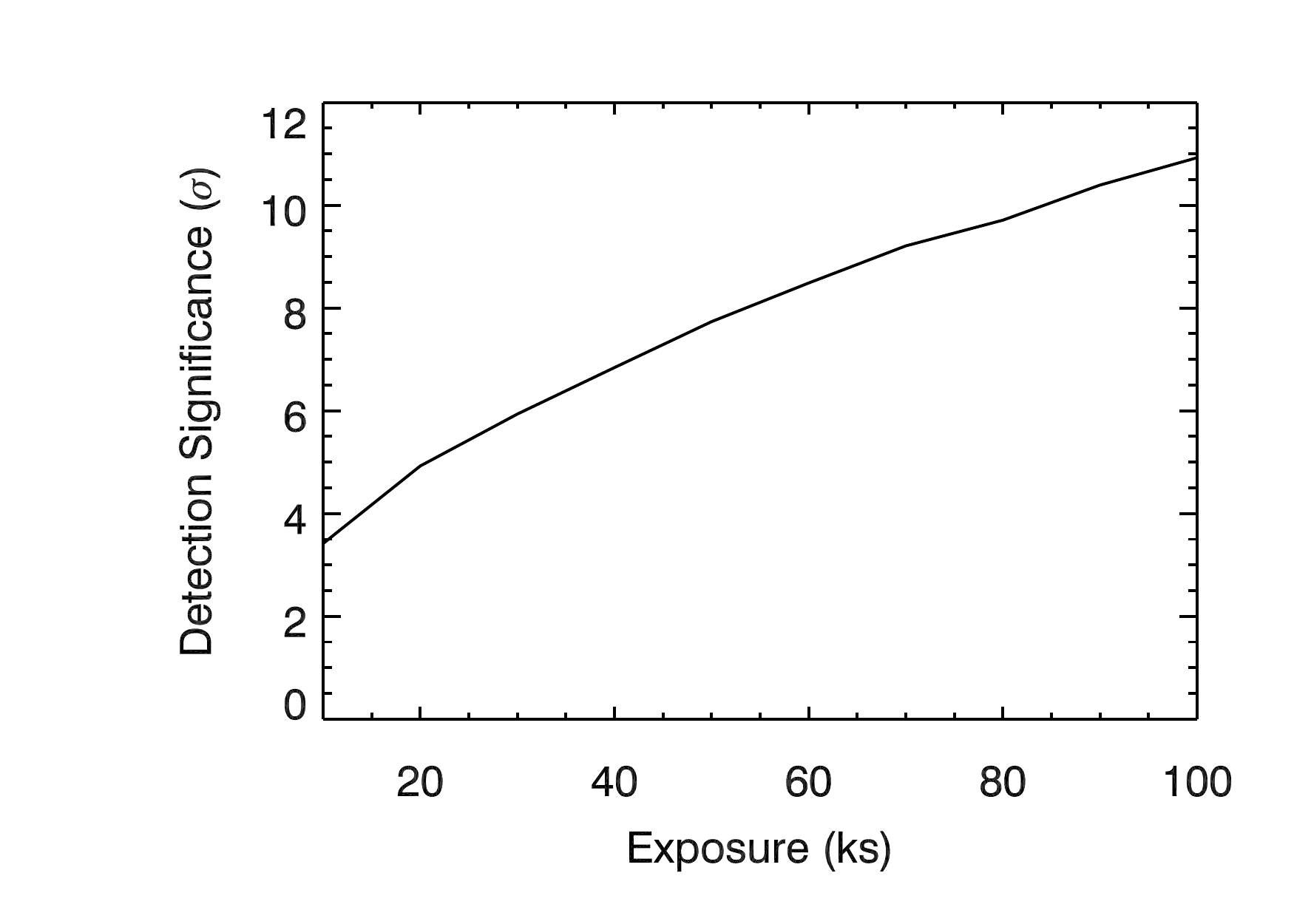}
    \caption{Simulations with \extp showing the
    significance of a detection of a PF$\sim$70\% (sufficient to prove QED
    effects) as a function of the exposure. Left: X-ray flux
    $\sim10^{-11}$\,erg\,cm$^{-2}$\,s$^{-1}$, as typical of the outburst 
onset;
    Right: flux $\sim5\times10^{-13}$\,erg\,cm$^{-2}$\,s$^{-1}$, as
    representative of the quiescent state after decay. }
\label{fig:qed5}
\end{figure*}

Simulations (based on parameters for XTE~J1810$-$197 and
CXO~J164710.2$−$455216, see \cite{ref90abcd,ref90abcde}) show that in order to
observe a PF$\sim$70\% and variation of the PA$\sim$10 deg, at the onset of the
outburst (with a maximum flux $\sim10^{-11}$\,erg\,cm$^{-2}$\,s$^{-1}$), the
observation time required by \extp is $\sim3$\,ks and $\sim20$\,ks respectively.
At the end of the X-ray flux decay, when the TM is in quiescence (minimum flux
$\sim5\times10^{-13}$\,erg\,cm$^{-2}$\,s$^{-1}$), the observation time required
to detect the same PF and PA by \extp is $\sim60$\,ks and $\sim380$\,ks,
respectively. Examples are reported in Fig.~\ref{fig:qed5} for the PF. 

Comparatively short exposures enabled by large effective area of the
\extp also make it the only instrument able to carry out systematic studies of
QED effects in multiple sources. Indeed, \extp will have in place a monitoring
programme of TMs, and will therefore represent a revolutionary instrument,
capable of testing QED effects in TMs systematically, with short observations,
in at least 1-2 sources per year.


\begin{figure*}[ht!]
\Centering
 \includegraphics[width=1.0\textwidth]{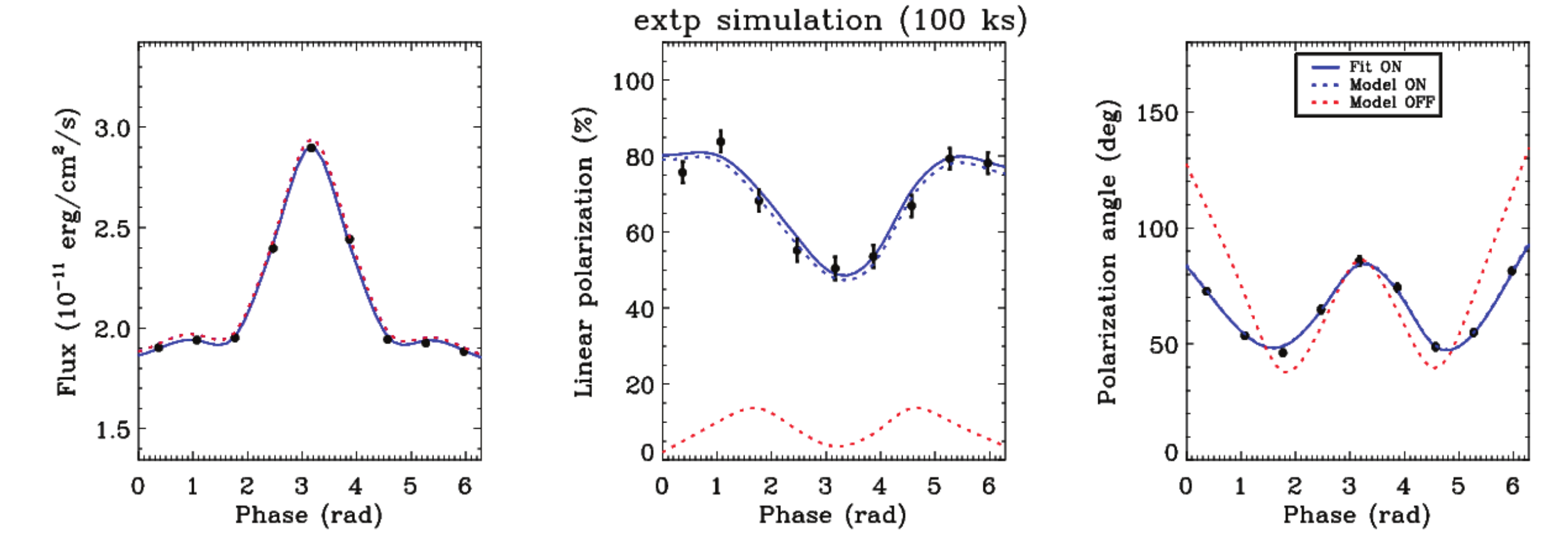}
    \caption{Light curve, degree and angle of polarization as computed
    according to the ``twisted magnetosphere'' model with $\Delta\phi_{N-S} =
    0.5$ rad, $\beta= 0.34$, $\chi= 90^{\circ}$ and $\xi = 60^{\circ}$ for a source
    with properties similar to those of the AXP 1RXS J170849.0-400910. Data
    points (filled circles with error bars) are generated assuming a 100 ks
    observation of the source and are drawn from the model shown by the blue
    dotted line; in both cases QED effects are fully taken into account (``ON''
    case). The red dotted line represents the same model but without vacuum
    birefringence (``OFF'' case). The simultaneous fit (solid blue line) of the
    ``ON'' data with the ``ON'' model gives a reduced $\chi^2=1.14$ while that
    of ``ON'' data with the ``OFF'' model is ruled out with high confidence.}%
\label{fig:rics_magnetar}
\end{figure*}

Fig.~\ref{fig:rics_magnetar} depicts a second example where we 
can use magnetar observations to probe QED effects, and to verify the 
nature of the source, in particular the 
presence of a strong magnetic field in the emission region. 
The leading model for $\sim1-10$\,keV X-ray emission from
magnetars is resonant inverse Compton scattering (RCS) of photons
from the surface by charges carried in currents through the
magnetosphere. In conjunction with the determination of the flux
and flux variation, models of this emission also predict the
change in the polarization of the high-energy magnetospheric
emission with the spin of the neutron star, as calculated by
\cite{ref91}. Fig.~\ref{fig:rics_magnetar} shows
the simulated \extp data for a bright persistent magnetar source
with properties similar to those of the AXP 1RXS J170849.0-400910.
Simulated data were obtained from the RCS model which best fits
the 0.5-10 keV spectrum of the source, properly accounting for QED
effects, and assuming a 100\,ks exposure time. The mock data were
then fitted with RCS models, either accounting for or disregarding QED
effects. As expected, the pulse profile is well fitted in both
cases (left panel), but the phase-dependence of the polarization
fraction and angle (middle and right panels) are recovered only by
QED-on models (middle and right panels). It is clear that, depending
on how well the underlying model can be constrained by other 
observations, the QED-off case may be excluded to a large degree
of confidence by such an observation with \extp.

Assuming discovery of the QED effect with \extp, there are two natural
directions in which to proceed.  One is to use the QED effect to probe the
other physical processes occurring in the source.  The second is to
measure the strength of the QED effect, effectively measuring the
index of refraction of the magnetized vacuum.  The left panel of
Fig.~\ref{fig:surface_magnetar} 
demonstrate this first avenue.  In Fig.~\ref{fig:surface_magnetar}, we
see how measurements of the polarized fraction of the surface emission
at photon energies of few keV could provide an estimate of the radius
of the neutron star (or more precisely $R_\infty=R [ 1-2GM/(c^2 R)
]^{-1/2}$).  Although the figure shows the surface field, the strength
of the QED effect depends on the strength of the dipole moment of the
neutron star, which can be estimated, in principle, from its spin down,
independent of the assumed radius of the star.  For this to work, we
would have to observe a more weakly magnetized neutron star that still
exhibits thermal emission from the surface up to a few keV, where \extp
is sensitive, yet not so weak that one does not expect the thermal
emission to be polarized at the surface ({\em i.e.} accreting
millisecond pulsars would probably not work, but accreting X-ray
pulsars might; however, in this case it might be difficult to obtain a
reliable estimate of the magnetic field strength, if CRSFs are not 
observed). 

The right panel of Fig.~\ref{fig:4u_herx1} depicts the polarized fraction as
function of energy averaged over phase for the accreting X-ray pulsar Her X-1.
We use the observed cyclotron resonance to constrain the magnetic field near
the surface of the star and we use the spectral models of \citep{ref63} and
\citep{ref88b}. Because the emission comes from only a small fraction of the
surface of the star, the effect of QED birefringence is to {\em reduce} the
observed polarized fraction. Furthermore, the effect of QED is not saturated in
this case, so with modeling one can determine the strength of the birefringence
and possibly constrain the radius of the star as well.

\subsection{Black Holes and White Dwarfs}
The theoretical treatment of the effects of vacuum polarization 
for black holes and white dwarfs is much less mature, but these
objects also could provide exciting probes of QED, and vice versa. 
Black hole accretion disks are expected to generate a 
magnetic field much weaker than that of neutron stars, so one might think 
that the effect of vacuum birefringence on X-ray polarization 
would be small. However, how strongly the birefringence affects 
the polarization of the photon traveling in the magnetized vacuum 
depends not only on the strength of the magnetic field itself,
but also on how long the photon travels in the strong 
magnetic field.  Preliminary calculations \cite{ref89} indicate that the vacuum
polarization becomes important around $5-10$~keV, right in the
middle of the \extp sensitivity band. 

\begin{figure*}[ht!]
\Centering
\includegraphics[width=0.4\textwidth]{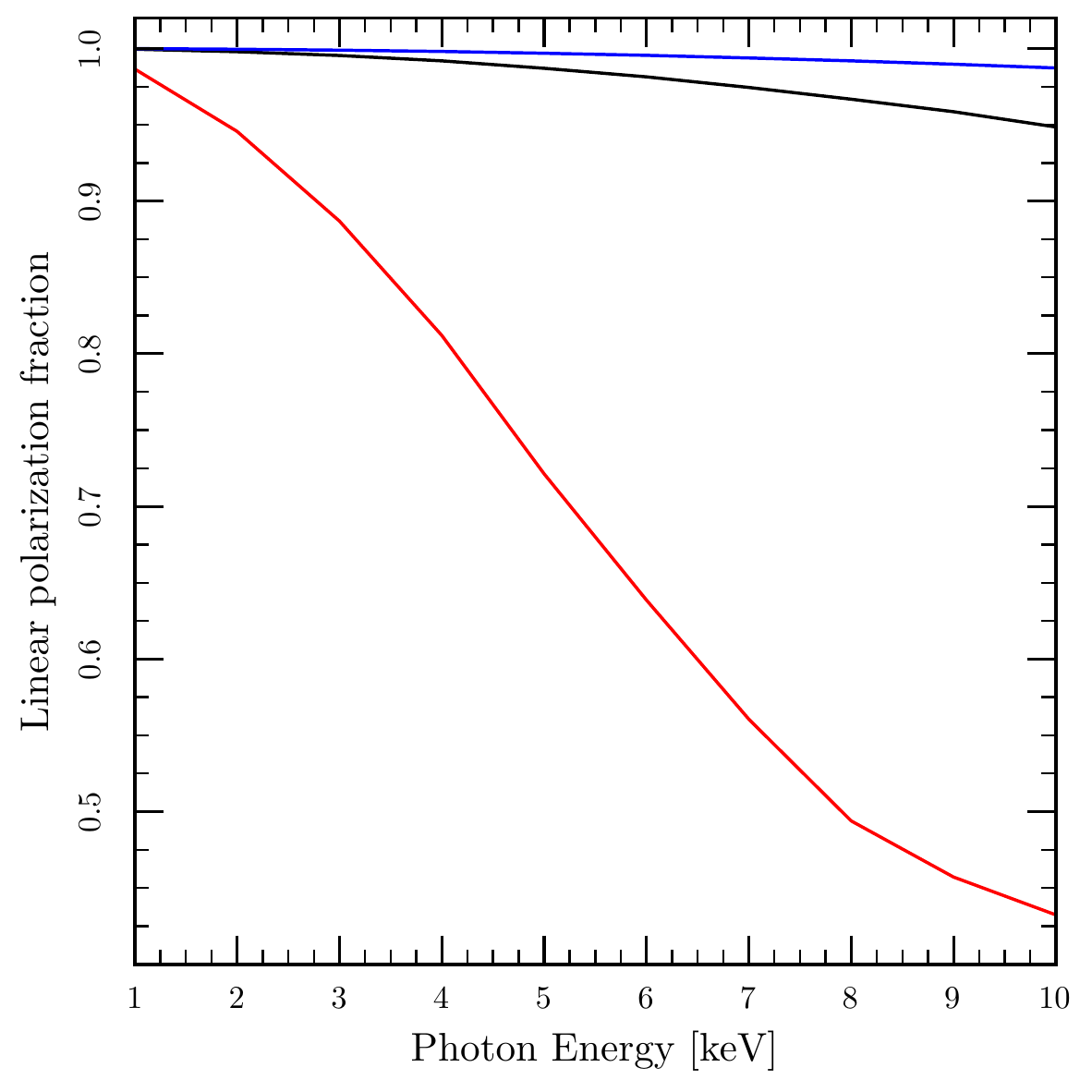}
\includegraphics[width=0.4\textwidth]{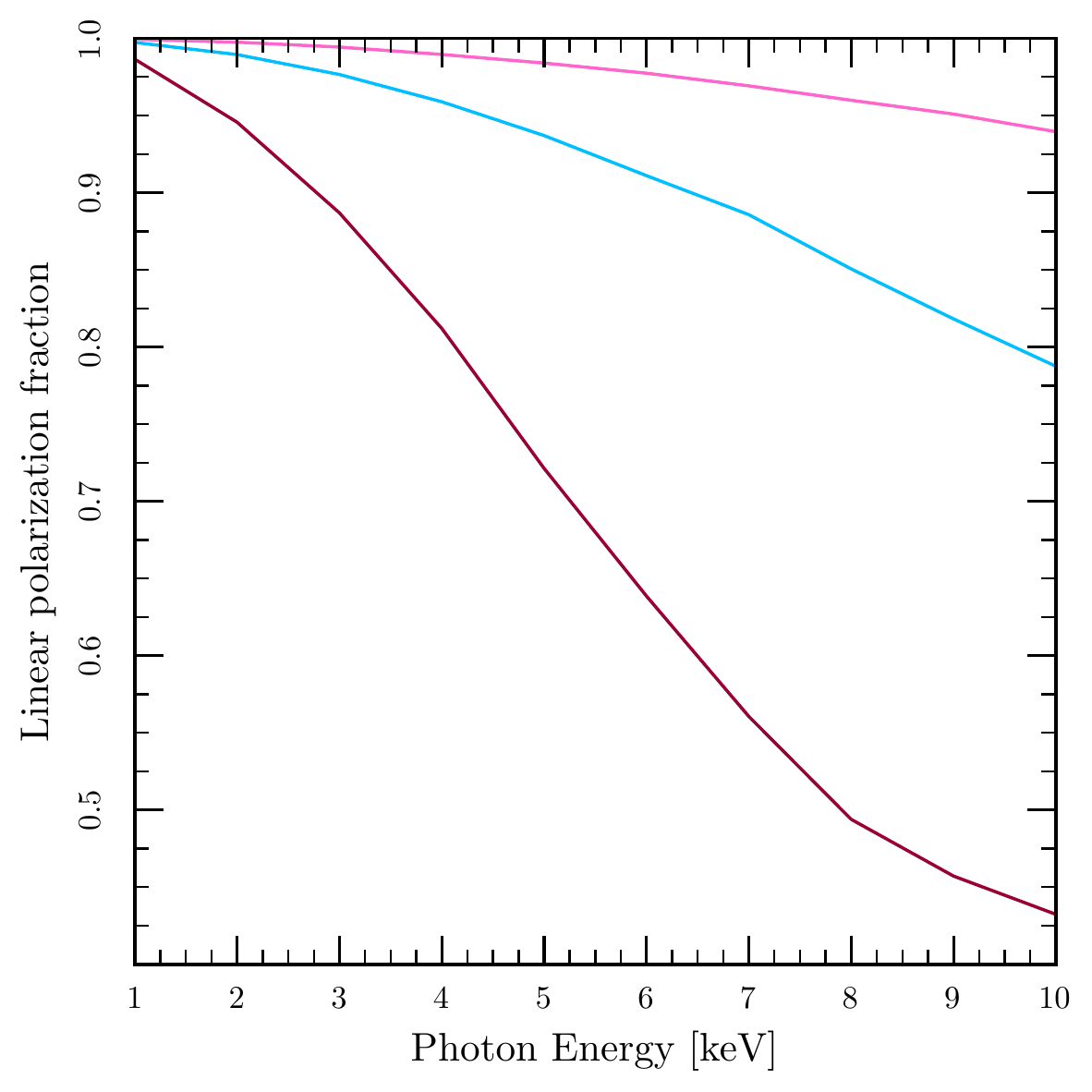}
\caption{Observed polarization fraction against photon energy for an
initial polarization fraction of one. Left: Zero angular 
momentum photons (black solid line), maximum prograde angular
momentum photons (highly blue-shifted, blue solid line) and 
maximum retrograde angular momentum photons (highly red-shifted,
red solid line), coming from the ISCO of a black hole with
$a_\star=0.9$. Right: Maximum retrograde angular momentum
photons for $a_\star=0.5$ (pink line), 0.7 (light blue line) 
and 0.9 (dark crimson line).}%
\label{fig:bh}
\end{figure*}

\cite{ref89} analyzed the effect of a chaotic magnetic field in 
the disk plane, for nearly edge-on observations of the disk. The 
chaotic field, in tandem with QED birefringence, depolarizes the 
radiation, as shown in Fig.~\ref{fig:bh}. \cite{ref89} performed 
Monte-Carlo simulations of 6,000 photons emitted with the same 
energy and angular momentum from the innermost stable circular
orbit (ISCO) of the accretion disk of a rotating black hole, 
calculating the evolution of the polarization as each photon 
travels through the magnetosphere. The same simulation was 
performed for three different angular momenta of the photon 
(zero, maximum retrograde and maximum prograde), for ten 
different energies (from 1 to 10 keV), and for three different 
spins of the black hole ($a_\star = a/M = J/cM^2 = 0.5$, 0,7 and 
0.9). The results are shown in Fig.~\ref{fig:bh}. As we can see from 
the left panel, the QED effect can change the observed 
polarization, especially for red-shifted photons at the high end 
of \extp's energy range, where the observed polarization can be
reduced up to 60\% for a black hole rotating at 
90\% the critical velocity. The effect is larger for rapidly rotating 
black holes, for which red-shifted photons perform many 
rotations around the hole before leaving the magnetosphere (see 
the right panel of Fig.~\ref{fig:bh}). These results are 
independent of the mass of the black hole.

This analysis shows that QED must be taken into account when modeling the
X-ray polarization from both stellar-mass and supermassive black hole 
accretion disks. In particular, it will be crucial in the understanding of 
QPOs, in which much of the variability in the flux results from
our receiving alternately red-shifted and blue-shifted photons at the
telescope. All the results in \cite{ref89} are obtained for the minimum
magnetic field needed for accretion to operate in the disk. A larger magnetic
field would cause a higher depolarization at all energies. The observation 
of
the X-ray polarization from black hole accretion disks would be, if 
properly
modeled including QED, the first probe of black hole magnetic fields and 
the
first direct way to study the role of magnetic fields in astrophysical
accretion generally.

The potential of magnetized accreting white dwarfs for polarization
measurements in the X-rays is tantalizing but completely unexplored. 
\cite{ref88}
argued that the polarized radiation should travel adiabatically near 
magnetized
white dwarfs. These magnetized systems are, of course, known as polars because
of the strong circular and linear polarization that they exhibit in the
optical. These optical observations, where it is likely that plasma effects
dominate, can give crucial information on the strength and geometry of the
magnetic field near the white dwarfs. Furthermore, basic quantities like the
radius of the white dwarf are well constrained theoretically and
observationally. With these data in hand, these objects become ideal
laboratories to probe QED, and in particular the more weakly magnetized white
dwarfs (the intermediate polars or DQ~Her stars) may offer the possibility
of measuring the index of refraction through observations of 
polarized X-ray emission.

%%%%%%%%%%%%%%%%%%%%%%%%%%%%%%%%%%%%%%%%%%%%%%%%%%%%%%%
Acknowledgements.
This paper is an initiative of the \extp's Science Working Group on Strong
Magnetism, whose members are representatives of the astronomical community at
large with a scientific interest in pursuing the successful implementation of
\extp. The paper was primaly written by by Andrea Santangelo, Silvia Zane, Hua Feng and
Renxin Xu. Major contributions by Andrea Santangelo, Victor Doroshenko, Mauro Orlandini (3.1-3.4); Renxin Xu, Zhaosheng Li, Lin Lin (2.2., 2.3); Hao Tong (2.5); Silvia Zane
(2.1, 2.2, 2.3.4, 2.4, 3.5, 4.1. 4.2); Nanda Rea, Paolo Esposito and Francesco Coti Zelati
(2.2.2); Gianluca Israel and Daniela Huppenkoten (2.3.4); Roberto Turolla and Roberto Taverna
(2.2.2., 4.2); Denis González-Caniulef (4.2); Roberto Mignani (3.5); Jeremy Heyl and Ilaria Caiazzo (4).
Contributions were edited by Andrea Santangelo, Silvia Zane and Victor Doroshenko. Other
co-authors provided valuable inputs to refine the paper.
Financial contribution from the agreement between the Italian Space Agency and
the Istituto Nazionale di Astrofisica ASI-INAF n.2017-14-H.O is acknowledged.
This work is also partially supported by the \textsl{Bundesministerium f\"{u}r
Wirtschaft und Technologie} through the \textsl{Deutsches Zentrum f\"{u}r Luft-
und Raumfahrt e.V. (DLR)} under the grant number FKZ 50 OO 1701.
%%%%%%%%%%%%%%%%%%%%%%%%%%%%%%%%%%%%%%%%%%%%%%%%%%%%%%%
%\Acknowledgements{This work was supported by the XXXX (Grant Nos. 6XXX) and CAS Interdisciplinary Project (Grant %XXX).}
\InterestConflict{The authors declare that they have no conflict of interest.}

%%%%%%%%%%%%%%%%%%%%%%%%%%%%%%%%%%%%%%%%%%%%%%%%%%%%%%%
%%% Conflict of interest. ????????????
%%%%%%%%%%%%%%%%%%%%%%%%%%%%%%%%%%%%%%%%%%%%%%%%%%%%%%%
%\InterestConflict{The authors declare that they have no conflict of interest.}

%%%%%%%%%%%%%%%%%%%%%%%%%%%%%%%%%%%%%%%%%%%%%%%%%%%%%%%
%%% Supplements. ????????, ????
%%%%%%%%%%%%%%%%%%%%%%%%%%%%%%%%%%%%%%%%%%%%%%%%%%%%%%%
%\Supplements{}

%%%%%%%%%%%%%%%%%%%%%%%%%%%%%%%%%%%%%%%%%%%%%%%%%%%%%%%
%%% Reference section. ??????
%%% citation in the content using "some words~\cite{1,2}".
%%% ~ is needed to make the reference number is on the same line with the word before it.
%%%%%%%%%%%%%%%%%%%%%%%%%%%%%%%%%%%%%%%%%%%%%%%%%%%%%%%

%\bibliographystyle{jwaabib}
%\bibliography{accrPulsars,magnetar,heyl_qed,sismo}

%%%%%%%%%%%%%%%%%%%%%%%%%%%%%%%%%%%%%%%%%%%%%%%%%%%%%%%
%%% Appendix sections. ??????, ????
%%%%%%%%%%%%%%%%%%%%%%%%%%%%%%%%%%%%%%%%%%%%%%%%%%%%%%%

\end{multicols}
\end{document}